\documentclass{iopart}
\usepackage{graphicx} 
\usepackage{bm} 
\usepackage{longtable}
\usepackage{multirow}

\newcommand{\text}[1]{\textrm{ #1}}
\newcommand{\bra}[1]{\langle #1 |}
\newcommand{\ket}[1]{| #1 \rangle}

\newcommand{\matelement}[3]{\langle #1 | #2 | #3 \rangle}
\newcommand{\beq}{\begin{equation}}
\newcommand{\eeq}{\end{equation}}
\newcommand{\crea}[2]{\hat #1^{\dagger}_{#2}}
\newcommand{\anni}[2]{\hat #1_{#2}}
\newcommand{\creation}[2]{\hat #1^{\dagger}_{#2}}
\newcommand{\annihilation}[2]{\hat #1_{#2}}

\def\r{\mathbf{r}}

\begin{document}
  
\title{Entanglement creation in cold molecular gases using strong laser pulses}

\author{Felipe Herrera}
\ead{fherrera@purdue.edu}
\address{Department of Chemistry, Purdue University, West Lafayette, IN 47907, USA}
\address{Department of Chemistry and Chemical Biology, Harvard University, 12 Oxford St., Cambridge, MA 02138, USA}

\author{Sabre Kais}
\address{Department of Chemistry, Purdue University, West Lafayette, IN 47907, USA}

\author{K. Birgitta Whaley}
\address{Department of Chemistry, University of California, Berkeley, CA 94703, USA}


\begin{abstract}

While many-particle entanglement can be found in natural solids and strongly
interacting atomic and molecular gases, generating highly entangled states
between weakly interacting particles in a controlled and scalable way presents a significant
challenge. We describe here a one-step method to generate entanglement in a dilute gas of 
cold polar molecules. For molecules in optical traps
separated by a few micrometers, we show that maximally entangled states can be
created using the strong off-resonant pulses that are routinely used in molecular
alignment experiments. We show that the resulting alignment-mediated entanglement
can be detected by measuring laser-induced fluorescence with single-site resolution 
and that signatures of this molecular entanglement also appear in the
microwave absorption spectra of the molecular ensemble. We analyze the robustness of these
entangled molecular states with respect to intensity fluctuations of the trapping laser and discuss possible 
applications of the system for quantum information processing.
\end{abstract}

\maketitle

The concept of entanglement has evolved from being regarded as a perplexing and even undesirable consequence of quantum mechanics in the early studies by Schr\"{o}dinger \cite{Schrodinger:1935,Wheeler&Zurek-book} and Einstein \cite{EPR:1935}, to being now widely considered as a fundamental technological resource that can be harnessed in order to perform tasks that exceed the capabilities of classical systems \cite{Horodecki:2009review}. Besides its pioneering applications in secure communication protocols and quantum computing \cite{Nielsen&Chuang-book}, entanglement has also been found to be an important unifying concept in the 
analysis of magnetism \cite{Ghosh:2003,New-Kais1,New-Kais2,Amico:2008review}, electron correlations \cite{New-Kais3} and quantum phase transitions \cite{Osborne:2002,Osterloh:2002,Amico:2008review}. Many properties and applications of entanglement have been demonstrated using a variety of physical systems including photons \cite{Aspect:1981,Gisin:1998,Zeilinger:1998,Zhao:2004,Peng:2005}, trapped neutral atoms \cite{Mandel:2003,Bloch:2008,Urban:2009,Wilk:2010,Isenhower:2010}, trapped ions \cite{Turchette:1998,Haffner:2005,Blatt:2008,Jost:2009,Moehring:2009}, and hybrid architectures \cite{Blinov:2004,Fasel:2005}. Entanglement has also been shown to persist in macroscopic \cite{Berkley:2003,Yamamoto:2003,Steffen:2006,Lee:2011} and biological systems \cite{Engel:2007,Sarovar:2010}. Despite this significant progress, the theory of quantum entanglement and its technological implications are still far from being completely 
understood \cite{Horodecki:2009review}.  

Trapped neutral atoms are regarded as a promising platform for applications of quantum entanglement due their relatively long coherence times \cite{Bloch:2008}, which can exceed those of solid state and trapped ion architectures by orders of magnitude \cite{Ladd:2010}. 
Moreover, the sources of single-particle decoherence are well characterized in electromagnetic traps \cite{Bloch:2008}, and can be compensated using standard state transfer techniques \cite{Bergmann:1998}. In order to address individual atoms in an optical trap for coherent state manipulation, it is necessary to separate the particles from each other by a distance comparable to optical wavelengths \cite{Bloch:2005,Weitenberg:2011}. 
However, it is difficult to achieve entanglement between ground state atoms at such long distances, due to the short range nature of their mutual interaction.  It is nevertheless is possible to enhance interactions between atoms in optical traps by either controlling the interatomic distance \cite{Jaksch:1999, Duan:2003,Hayes:2007}, or exciting atoms to an internal state that supports long-range interactions \cite{Brennen:2000,Deutsch:2005,Jaksch:2000,Lukin:2001,Saffman:2005}. Using these methods, recent experiments have demonstrated the generation and characterization of entangled 
atomic states \cite{Mandel:2003, Anderlini:2007, Wilk:2010,Isenhower:2010}, which are the first steps towards the study of many-particle entanglement and the development of quantum technologies using optically trapped particles. 

Quantum entanglement can also be studied using trapped polar molecules \cite{Carr:2009}. Arrays of polar molecules can be prepared in optical lattices with full control over the internal states including the hyperfine structure \cite{Ospelkaus:2006,Ni:2008,Ospelkaus:2010-hyperfine,Chotia:2012}. Trapped molecules inherit the long coherence times of their atomic counterparts and the long-range dipole-dipole interaction between molecules offers a route for entanglement generation. Since the dipole moment of freely rotating molecules averages to zero, proposals for molecular entanglement creation have involved the application of DC electric fields to spatially orient the dipoles \cite{Yelin:2009}. One promising approach consists of placing the oriented dipoles in an ordered array using an optical lattice and performing entangling gate operations using microwave pulses, building on analogies with architectures for NMR quantum computation \cite{DeMille:2002,Wei:2011,Zhu:2013}. In order to overcome the complexity 
involved in 
controlling the ``always-on'' interaction between oriented dipoles, conditional transitions between weakly and strongly interacting states have also been proposed 
as a route to generation of intermolecular entanglement \cite{Yelin:2006,Charron:2007,Kuznetsova:2008}. This approach has recently been demonstrated experimentally for cold atoms \cite{Wilk:2010,Isenhower:2010}. Theoretical work has shown that entanglement can also be generated by coupling internal states with
collective motional states in strongly interacting molecular arrays \cite{Rabl:2007,Ortner:2011}, analogously to methods developed for trapped ions \cite{Soderberg:2010}. 
In addition to these approaches for the controlled generation of pairwise entanglement between molecules, many-particle entanglement is also expected to emerge in the pseudo-spin dynamics of an ensemble of polar molecules with tunable interactions \cite{Micheli:2006,Herrera:2010,Jesus:2010,Gorshkov:2011prl,Baranov:2012}.

In contrast with previous approaches for generation of entanglement between dipolar molecules, the scheme proposed here does not involve the use of DC electric fields. 
Instead, we introduce here a method for deterministic generation of entanglement 
that uses strong optical laser pulses far-detuned from any vibronic transition. We consider closed-shell polar molecules in their ground rovibrational state, with each molecule individually confined in an optical trap in order to suppress collisional losses. 
We show that a single off-resonant laser pulse can mediate the entanglement of weakly interacting polar molecules separated by up to several micrometers. The degree of entanglement and the timescale of the entanglement operation are shown to have a well-defined dependence on experimental parameters such as the pulse intensity and duration. The laser parameters considered in this work are consistent with the technology developed to study molecular alignment in thermal gases \cite{Friedrich:1995,Sakai:1999,Stapelfeldt:2003,Seideman:2005}. 
We note that entanglement of polar rigid rotors in strong laser fields has been considered before in the high-density regime \cite{Liao:2004,Liao:2006}, where the dipole-dipole interaction energy is comparable to the rotational constant. 
The approach presented here allows for the generation of laser-mediated entanglement of rotors in dilute gases for the first time.

The remainder of this paper is organized as follows.  Section \ref{sec:ac fields} reviews the rotational structure of closed-shell molecules in strong off-resonant optical fields. In Section \ref{sec:entanglement generation} we analyze the generation of entanglement between two distant polar molecules due to the action of a single off-resonant laser pulse. The dependence of the degree of entanglement on experimental parameters is discussed in detail. In section \ref{sec:entanglement quantification} we discuss two entanglement detection schemes, one based on Bell-type measurements for systems possessing single-molecule addressability and another scheme that employs microwave spectroscopy with only global addressing capability. In section \ref{sec:decoherence} we investigate the effects of motional decoherence and show that entanglement in optical traps can be robust against this type of noise. We close with a summary and conclusions in Section \ref{sec:conclusions}. 

\section{Molecules in far-detuned optical fields}
\label{sec:ac fields}

We consider closed-shell diatomic molecules in the vibrational and electronic ground state. The state of the molecules in the absence of external fields is represented by $\ket{N,M_N}$, which is an eigenstate of the rigid rotor Hamiltonian $\hat H_{\rm R} = B_{\rm e}\hat N^2$ and $\hat N_Z$, where $\hat N$ is the rotational angular momentum operator and $\hat N_Z$ its component along the space-fixed $Z$-axis. $B_{\rm e}$ is the rotational constant. The interaction of a molecule with a monochromatic electromagnetic field $\mathbf{E}(\r,t)=\frac{1}{2}\left[ \hat\epsilon E(t)e^{i\omega t} + c.c. \right]$ whose frequency $\omega$ is far-detuned from any vibronic resonance can be described by the time-independent effective Hamiltonian \cite{Seideman:2005}
\begin{equation}
 \hat H_{\rm AC} =  -\sum_{p,p'}\hat\alpha_{p,p'}E_{p}(\r)E^*_{p'}(\r),
\label{eq:ac space fixed}
\end{equation}
where $E_p(\r)$ is the space-fixed $p$-component of the positive-frequency field in the spherical basis and $\hat \alpha_{p,p'}$ is the 
molecular polarizability operator. 
For diatomic molecules in a linearly polarized field, transforming the polarizability operator to the rotating body-fixed frame allows Eq. (\ref{eq:ac space fixed}) to be rewritten as
\begin{equation}
  \hat H_{\rm{AC}} = -\frac{|E_0|^2}{4}\left\{\frac{1}{3}(\alpha_{\parallel}+2\alpha_{\perp})+\frac{2}{3}(\alpha_{\parallel}-\alpha_{\perp}) \mathcal{D}^{(2)}_{0,0}(\theta)\right\},
\label{eq:ac diatomic}
\end{equation}
where $\mathcal{D}^{(2)}_{0,0}=(3\cos^2\theta -1)/2$ is an element of the Wigner rotation matrix \cite{Zare}, $E_0$ is the field amplitude 
for the selected polarization and $\theta$ is the polar angle of the internuclear axis with respect to 
this. The polarizabilty tensor for diatomic molecules is parametrized by its parallel $\alpha_\parallel$ and perpendicular $\alpha_\perp$ components, 
with $\alpha_\parallel>\alpha_\perp$. 
The first term in Eq. (\ref{eq:ac diatomic}) leads to a state-independent shift of the rotational levels and the second term induces coherences between rotational states $\ket{NM_N}$, according to the selection rules $\Delta N=0,\pm 2$ and $\Delta M_N = 0$. Therefore the parity of rotational states in the presence of a far-detuned field is conserved.

\begin{figure}[t]
\begin{center}
 \includegraphics[width=0.70\textwidth]{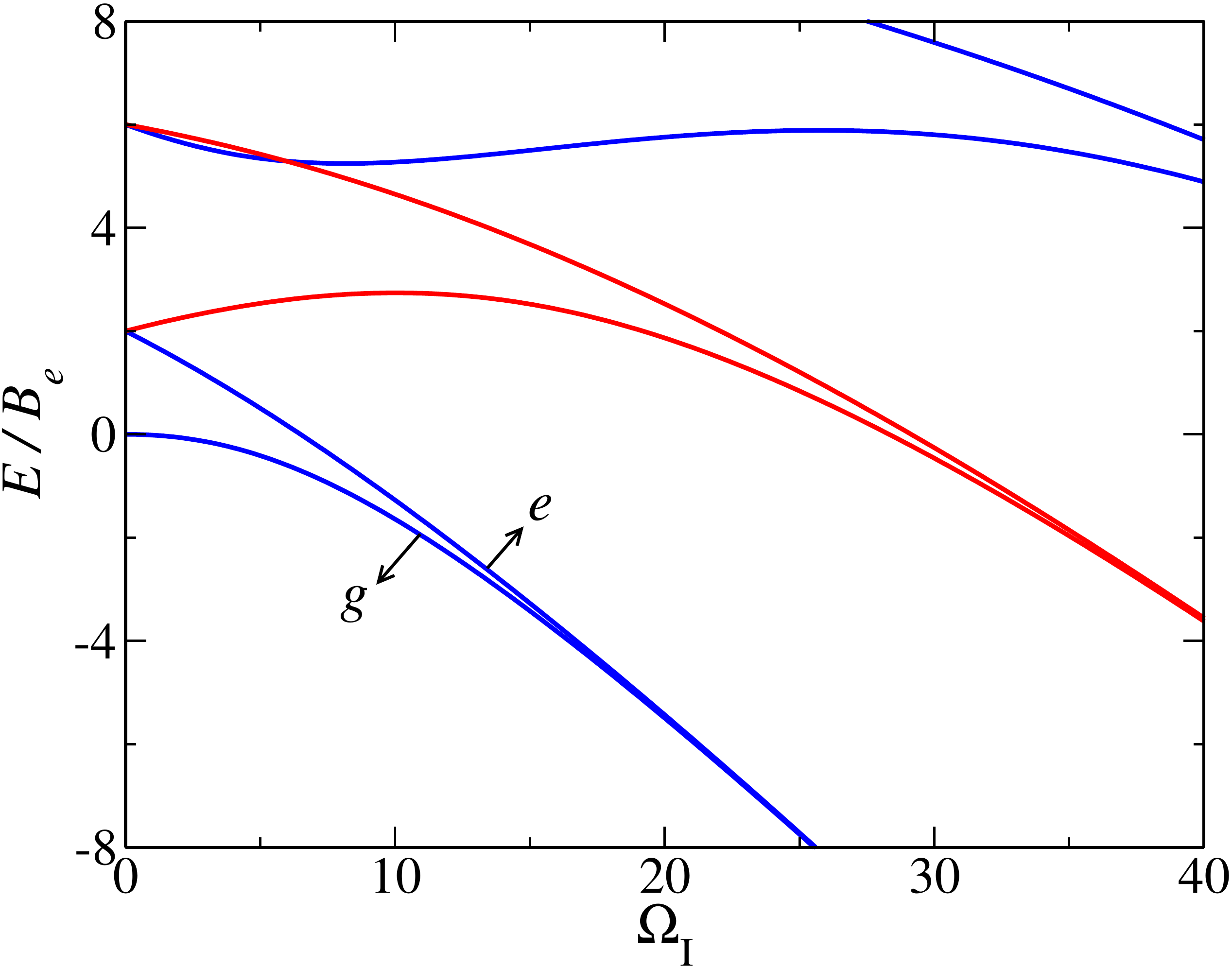}
\end{center} 
  \caption{Dimensionless rotational energy $E/B_{\rm e}$ of a molecule in the presence of a linearly-polarized CW far-detuned laser, as a function of the light-matter coupling strength $\Omega_{\rm{I}} = |E_0|^2\Delta\alpha/4B_{\rm e}$: (a) Energies of the first six states with $M_N=0$ (blue) and $|M_N|=1$ (red). The states of the lowest doublet $\ket{g}=\ket{\tilde 0,0}$ and $\ket{e}=\ket{\tilde 1,0}$ define a two-level subspace. 
  $B_{\rm e}$ is the rotational constant, $\Delta\alpha$ is the polarizability anisotropy, and $|E_0|^2 = I/2\epsilon_0c$, where $I$ is the intensity of the laser. The notation $\ket{\tilde N,M_N}$ indicates that the rotational quantum number $N$ is not conserved for $\Omega_{\rm I}\neq 0$. $M_N$ is the projection of the rotational angular momentum along the laser polarization.}
\label{fig:ac energies}
\end{figure}
Ignoring the state-independent light shift (which contributes with just an overall phase to the eigenstates) and expressing the energy in units of $B_{\rm e}$, the single-molecule Hamiltonian $\hat H = \hat H_{\rm R}+\hat H_{\rm AC}$ can then be written as 
\begin{equation}
 \hat H = \hat N^2 - \frac{2}{3}\Omega_{\rm I} \mathcal{D}^{(2)}_{0,0}(\theta),
\label{eq:dimless ac}
\end{equation}
where $\Omega_{\rm I} = {|E_0|^2(\alpha_{\parallel}-\alpha_{\perp})}/{4B_{\rm e}}$ 
is a dimensionless parameter that characterizes the strength of the light-matter interaction and is proportional to the field intensity $I_0=\frac{1}{2}c\epsilon_0|E_0|^2$. In Fig. \ref{fig:ac energies} we plot the lowest eigenvalues of $\hat H$ as a function of 
$\Omega_{\rm I}$. The figure shows that for intense fields $\Omega_{\rm I}\gg 10$, the energy spectrum consists of closely spaced doublets, as first discussed in Ref. \cite{Friedrich:1995}. 
The lowest doublet states $\ket{g}$ and $\ket{e}$ correlate adiabatically with the states $\ket{g}\equiv\ket{0,0}$ and $\ket{e}\equiv\ket{1,0}$ in the limit  $\Omega_{\rm I}\rightarrow0$.
Since the eigenstates of Hamiltonian in Eq. ($\ref{eq:dimless ac}$) have well-defined parity, 
the induced dipole moments $\bra{g}\mathbf{d}\ket{g}$ and $\bra{e}\mathbf{d}\ket{e}$ vanish, but the transition dipole moment $\bra{e}\mathbf{d}\ket{g}$ is finite for polar molecules, where $\mathbf{d}$ is the electric dipole operator.

The light-matter interaction term $\hat H_{\rm AC}$ in Eq. (\ref{eq:ac diatomic}) has been widely used to describe the alignment of polar and non-polar molecules in intense off-resonant fields \cite{Friedrich:1995, Bonin:1997,Seideman:2005}. 
From a classical point of view, the electric field of a strong off-resonant optical field polarizes the molecular charge distribution, inducing an instantaneous dipole moment. The field then exerts a torque on the rotating dipole that changes the angular momentum of the molecule, favouring the alignment of the dipole axis along the field polarization direction. However, the orientation of the dipole is not well-defined in AC electric fields. 
The degree of alignment for diatomic molecules is typically measured by the expectation value $\mathcal{A} = \langle\cos^2\theta\rangle$ \cite{Seideman:2005, Sakai:1999,Stapelfeldt:2003}, with $\theta$ defined in Eq. (\ref{eq:ac diatomic}). $\mathcal{A}$ is close to unity for aligned molecules. Adiabatic alignment in the presence of strong off-resonant laser pulses has been extensively studied both experimentally and theoretically \cite{Stapelfeldt:2003,Seideman:2005}. In adiabatic alignment experiments the laser pulse turn-on and turn-off times are long compared with the free rotational timescale $t_R\equiv \hbar/B_{\rm e}$. Under adiabatic conditions, the rotational motion of the molecules is described by the eigenstates 
of Eq. (\ref{eq:dimless ac}) with adiabatically varying values of $\Omega_{\rm I}(t)$.

In this work we consider molecules driven by strong off-resonant pulses that are adiabatic with respect to the rotational timescales, but not necessarily adiabatic with respect to longer timescales such as the dipole-dipole interaction time between distant molecules (see below).

\section{Dynamical entanglement generation using strong laser pulses}
\label{sec:entanglement generation}

We now consider the dipole-dipole interaction between polar molecules in the presence of a strong off-resonant laser. The single-molecule Hamiltonian $\hat H = \hat H_{\rm R}+\hat H_{\rm AC}$ is given in Eq. (\ref{eq:dimless ac}) with intensity-dependent eigenvalues shown in Fig. \ref{fig:ac energies}. 
Using the two-level single-molecule subspace $\mathcal{S}_1 = \left\{\ket{g},\ket{e}\right\}$ the dipole-dipole interaction operator
can be written as 
\begin{equation}
\hat V_{\rm dd} = \gamma(1-3\cos^2\Theta) U_{\rm dd}(R)\times\left\{\ket{g_1e_2}\bra{e_1g_2}+\ket{e_1e_2}\bra{g_1g_2}+\text{H.c.}\right\}, 
\label{eq:exchange coupling}
\end{equation}
where $\gamma = d^{-2}\matelement{e}{\hat d_0}{g}^2$ is a universal dimensionless parameter that depends on the external field strength and polarization, $U_{\rm dd}=d^2/R^3$ is the interaction energy scale, $R$ is the intermolecular distance, $\Theta$ is the polar angle of the intermolecular axis with respect to the laser polarization, $\hat d_0$ is the component of the electric dipole operator along the laser polarization and $d$ is the permanent dipole moment of the molecule. 
At distances such that $U_{\rm dd}/B_{\rm e}\ll 1$, the interaction operator $\hat V_{\rm dd}$ does not mix the states $\ket{g}$ and $\ket{e}$ with higher field-dressed rotational states.

The two-molecule Hamiltonian matrix $\mathcal{H} = \hat H_1+\hat H_2+\hat V_{\rm dd}$ in the subspace $\mathcal{S}_2=\left\{\ket{g_1g_2},\ket{g_1,e_2},\ket{e_1g_2},\ket{e_1,e_2}\right\}$ can be written in two equivalent forms (up to a constant energy shift) as
\begin{eqnarray}
 \mathcal{H} &=& \varepsilon_{\rm e}\left(\crea{c}{1}{×}\annihilation{c}{1}+\crea{c}{2}{×}\annihilation{c}{2}\right)+J_{12}\left(\creation{c}{1}+\annihilation{c}{1}\right)\left(\creation{c}{2}+\annihilation{c}{2}\right)\nonumber\\
  &=&\frac{\varepsilon_{\rm e}}{2}(\sigma_Z^1+\sigma_Z^2)+J_{12}\sigma_X^1\sigma_X^2,
  \label{eq:H second-quantized}
\end{eqnarray}
where the operator $\creation{c}{i}=\ket{e_i}\bra{g_i}$ creates a rotational excitation on the $i$-th molecule, 
with the states $\ket{g_i}$ and $\ket{e_i}$ equivalently represented by eigenstates of $\sigma_Z^i$ with eigenvalues $-1,+1$, respectively, where 
$\sigma_\alpha^i$ ($\alpha = X,Y,Z$) is a spin-1/2 Pauli matrix.
 $J_{12} \equiv \bra{e_1g_2}\hat V_{\rm dd}\ket{g_1e_2}=\bra{e_1e_2}\hat V_{\rm dd}\ket{g_1g_2}$ is the exchange coupling energy, and $\varepsilon_{\rm e}$ is the splitting of the lowest doublet in Fig. \ref{fig:ac energies}. 
The eigenstates of $\mathcal{H}$ involving the single excitation sector are the symmetric and antisymmetric Bell states $ \ket{\Psi_\pm}=2^{-1/2}\left\{\ket{g_1e_2}\pm\ket{e_1g_2}\right\}$
with the eigenvalues $E_\pm=\varepsilon_{\rm e}\pm J_{12}$. The ground and highest excited states can be written as
\begin{equation}
\begin{array}{lcr}
 \ket{\Phi_{-}(\alpha)}&=&\cos\alpha\,\ket{g_1g_2}-\sin\alpha\,\ket{e_1e_2}\\
  \ket{\Phi_+(\alpha)}&=&\sin\alpha\,\ket{g_1g_2}+\cos\alpha\,\ket{e_1e_2}
\end{array},
\label{eq:adiabatic states}
\end{equation}
with eigenvalues $E_{\pm}=\varepsilon_{\rm e}\pm K$, where $K=\sqrt{\varepsilon_{\rm e}^2+J_{12}^2}$. The states $\ket{\Phi_{\pm}(\alpha)}$ are linear combinations of the remaining Bell states $\ket{\Phi^\pm} = 2^{-1/2}\left\{\ket{g_1g_2}\pm\ket{e_1e_2}\right\}$. The mixing angle $\alpha$ is defined by $\tan(2\alpha)=J_{12}/\varepsilon_{\rm e}$. The states $\ket{\Phi_\pm(\alpha)}$ are separable in the limits $\alpha \rightarrow 0$ and $\alpha\rightarrow \pm\infty$.  The ground state of the system is $\ket{\Phi_-(\alpha)}$ for all values of $\alpha$. 

Since the eigenstates of this two-molecule Hamiltonian are entangled for any finite value of the ratio $J_{12}/\varepsilon_{\rm e}$, 
we may consider the possibility of tuning the degree of entanglement by manipulating the transition energy $\varepsilon_{\rm e}$ with a strong off-resonant field. This corresponds to varying the effective magnetic field $h = \varepsilon_{\rm e}/2$ for the spin chain Hamiltonian in Eq. (\ref{eq:H second-quantized}). The possibility of preparing the states $\ket{\Phi_\pm(\alpha)}$ in Eq. (\ref{eq:adiabatic states}) using strong continuous-wave (CW) off-resonant laser fields was first pointed out in Ref. \cite{Lemeshko:2012}. However, 
since in practice the achievable intensity of CW lasers is limited, we consider here an alternative dynamical preparation of molecular entanglement using pulsed lasers.

Polar molecules can be prepared in the rovibrational ground state $\ket{g}$ inside an optical trap \cite{Carr:2009}. 
A strong linearly polarized off-resonant field can then be used to bring the energy of the excited state $\ket{e}$ close to degeneracy with the ground state $\ket{g}$ by adiabatically following the energy level diagram in Fig. \ref{fig:ac energies}. In the presence of a laser pulse, both the dipolar coupling $J_{12}(t)$ and the excitation energy $\varepsilon_{\rm e}(t)$ become time-dependent. We take the initial two-molecule wavefunction as $\ket{\Psi(0)}=\ket{g_1g_2}$. For this initial condition the state evolution is determined by the Hamiltonian sub-block
\begin{equation}
 \mathcal{H} =\left(\begin{array}{cc} 	
 0	&   J_{12}(t)	\\
 J_{12}(t)		&2\varepsilon_{e}(t) \\
\end{array} \right),
\label{eq:two-level matrix}
\end{equation}
with no participation of the single-excitation manifold since the Hamiltonian in Eq. (\ref{eq:H second-quantized}) is block-diagonal. The state of the system is described by a superposition of the form 
\begin{equation}
 \ket{\Phi(t)}=a(t)\ket{g_1g_2}+b(t)\ket{e_1e_2}.
\label{eq:final state}
\end{equation}
Expressing the energy in units of the rotational constant $B_{\rm e}$ and time in units of $t_{\rm R}=\hbar/B_{\rm e}$, we can write the equations of motion $i\dot a(\tau)=J(\tau)b(\tau)$ and $i\dot b(\tau)=J(\tau)a(\tau) + 2E(\tau)b(\tau)$, which we integrate numerically using a standard Runge-Kutta-Fehlberg method \cite{Pozrikidis-book}. We have defined here the dimensionless energies $J=J_{12}/B_{\rm e}$, $E = \epsilon_e/B_{\rm e}$, and time $\tau =t/t_{\rm R}$. The dipole-dipole interaction timescale $t_{\rm dd}=\hbar/U_{\rm dd}$ depends on the intermolecular distance.
The ratio between the rotational and interaction timescales $t_{\rm dd}/ t_{\rm R}$ is larger than unity for distances larger than the characteristic dipole radius (in atomic units)
\begin{equation}
 R_0=\left(d^2/B_{\rm e}\right)^{1/3}.
\label{eq:R0}
\end{equation}

We solve the time-dependent Schr\"{o}dinger equation by evaluating the energies $E(t)$ and $J(t)$ at each time step using an intensity parameter of the form $\Omega_{\rm I}(t) = [f(t)]^2\Omega_0$, for a Gaussian electric field envelope $f(t)=\rme^{-(t/t_0)^2}$. We take $t_0\gg t_{\rm R}$ to ensure adiabaticity with respect to the rotational motion. 
Under this condition we may extract $E(t)$ from Fig. \ref{fig:ac energies}. The exchange energy $J_{12}(t)$ is evaluated using the instantaneous eigenstates $\ket{g(t)}$ and $\ket{e(t)}$ of the single-molecule Hamiltonian in Eq. (\ref{eq:dimless ac}). The parameter $\gamma$ varies in the range $1/3 \leq \gamma\leq 1$ as a function of $\Omega_{\rm I}$, increasing monotonically from its lower bound at $\Omega_{\rm I} =0$ and reaching unity asymptotically as $\Omega_{\rm I}$ increases. The presence of a weak DC electric field in addition to the time-dependent laser field significantly changes this simple behaviour.  We discuss the effect of a DC field in detail in \ref{sec:dc fields}. In the following we shall consider the evolution of the system in the absence of DC electric fields.

\begin{figure}[t]
\begin{center}
\includegraphics[width=0.70\textwidth]{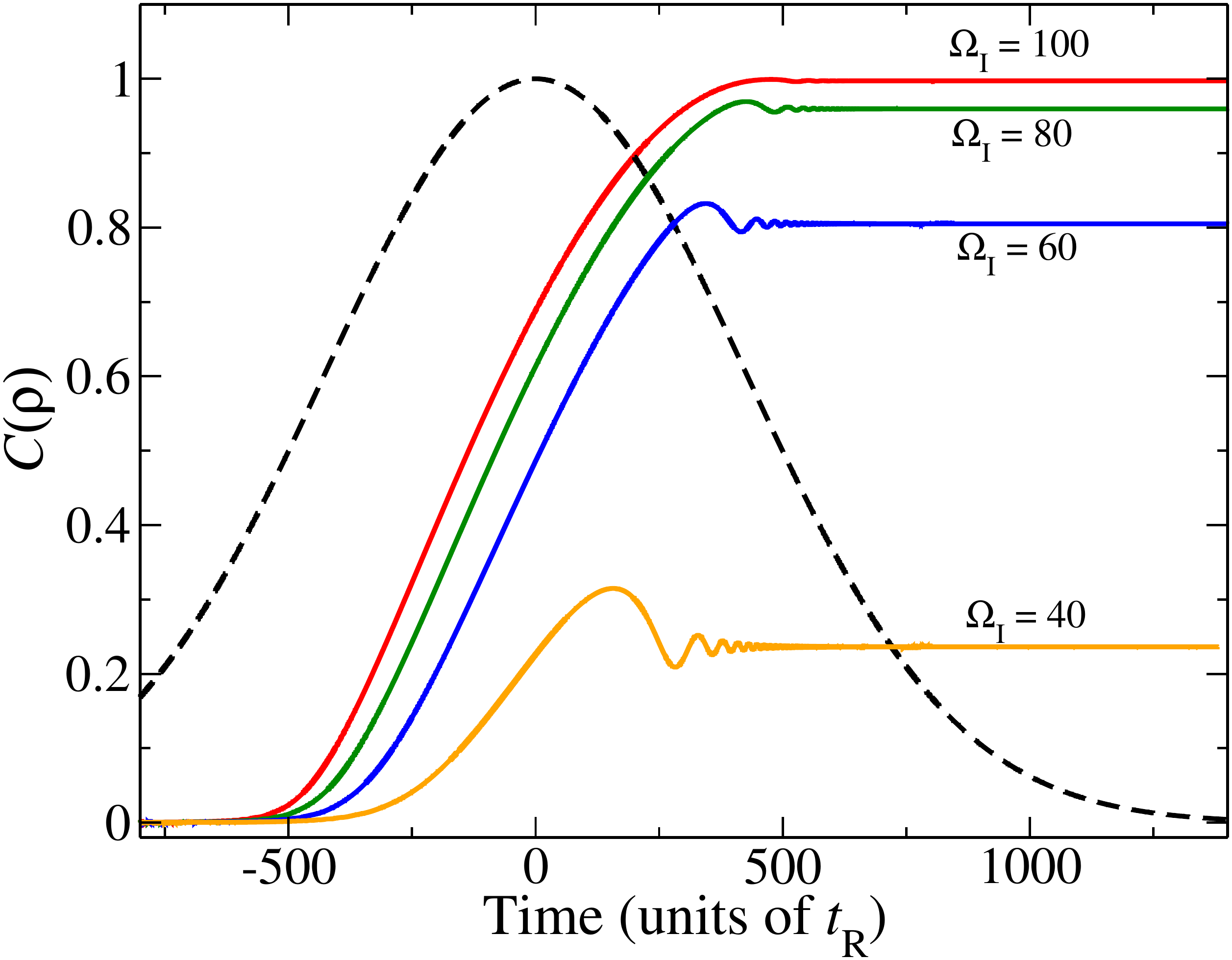}
\caption{ Evolution of the two-molecule concurrence $C(\rho)$ under the action of a Gaussian off-resonant laser pulse with intensity profile $\Omega_{\rm I}(t) = f^2(t)\Omega_0$, centered at $t = 0$. The intermolecular distance is $R = 10\,R_0$ and the pulsewidth $\tau_{\rm p}=t_{\rm dd}=10^3 t_{\rm R}$. Curves are labeled according to the value of the peak intensity $\Omega_0$. The dashed line shows the envelope function of the pulse $f(t)$. $t_{\rm dd}$ is the dipole-dipole interaction time and $t_{\rm R}=\hbar/B_{\rm e}$ is the rotational timescale.}
\label{fig:evolution}
\end{center}
\end{figure}

\subsection{Tuning entanglement with a single laser pulse}

We consider here pulses that are non-adiabatic with respect to the interaction timescale $t_{\rm dd}=(R/R_0)^3\,t_{\rm R}$. For a laser pulse that is adiabatic with respect to both $t_{\rm R}$ and $t_{\rm dd}$, an initial separable two-particle state would simply acquire a dynamical phase after the pulse is over and no net entanglement would be created in the system.

We define the entanglement radius $R_{\rm e}$ as the intermolecular separation at which the dipole-dipole interaction energy $U_{\rm dd}$ is equal to the energy of the transition $\ket{g_1g_2}\rightarrow\ket{e_1e_2}$, i.e., 
\begin{equation}
 R_{\rm e}= \left(d^2/2\varepsilon_{\rm e}\right)^{1/3}.
\label{eq:entanglement radius}
\end{equation}
\noindent
For two molecules within this radius, mixing of the states $\ket{g_1g_2}$ and $\ket{e_1e_2}$ is energetically allowed in the presence of a strong laser pulse. 
In the absence of DC electric fields, on account of the exponentially decreasing splitting of the doublet states as a function of the intensity parameter $\Omega_{\rm I}$, the entanglement radius $R_{\rm e}$ increases exponentially with  $\Omega_{\rm I}$. 
For concreteness, the value $\Omega_{\rm I}= 300$ corresponds to $R_{\rm e}\approx 3000 \,R_0$, which corresponds to distances of several micrometers between molecules (see \ref{sec:dc fields}). 

\begin{figure}[t]
\begin{center}
\begin{center}
\includegraphics[width=0.70\textwidth]{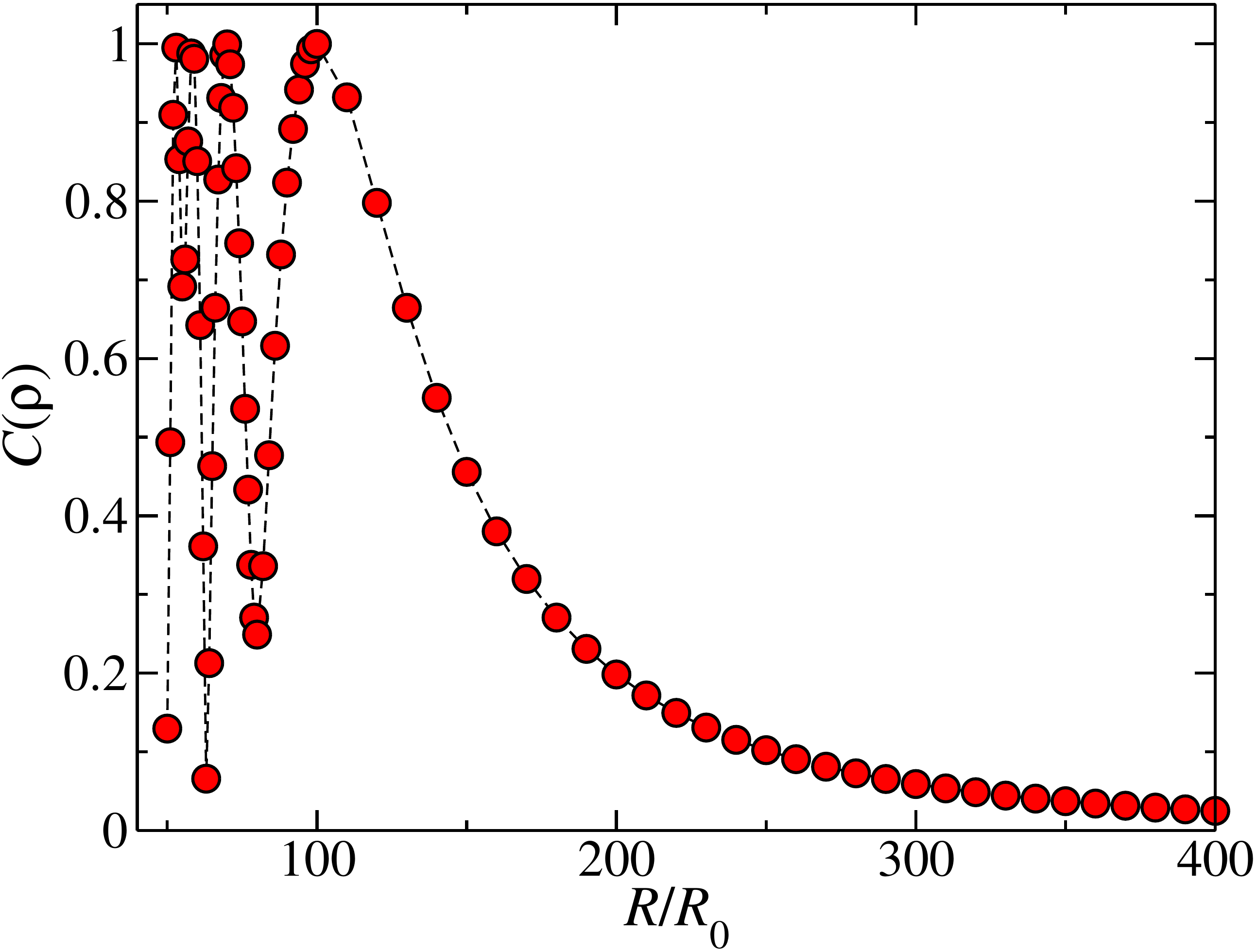}
\end{center}
\caption{Asymptotic two-molecule concurrence $C(\rho)$ as a function of the intermolecular distance $R$ (in units of $R_0$), long after the action of a Gaussian off-resonant laser pulse. For the distance $R=100\; R_0$, we choose the pulsewidth $\tau_{\rm p}=10^6\;t_{\rm R}$ (FWHM) and peak intensity $\Omega_0 = 270$, to obtain a maximally entangled state with $C(\rho)=1$. $t_{\rm dd}/t_{\rm R} = (R/R_0)^3$ is the dipole-dipole interaction time in units of the rotational timescale $t_{\rm R}=\hbar/B_{\rm e}$.}
\label{fig:distance dependence}
\end{center}
\end{figure}

Let us consider a pair of polar molecules separated by a distance $R_0\ll R < R_e$, where both molecules are initially in their rotational ground states, i.e., $\ket{\Psi(0)}=\ket{g_1g_2}$. The evolution of this system in the presence of a single Gaussian laser pulse
is given by Eq. (\ref{eq:final state}) and depends on three independent parameters: the intermolecular distance $R$, the pulse peak intensity $\Omega_0$, and the pulsewidth $\tau_{\rm p}$ (FWHM). We use the binary concurrence $C(\rho) = 2|ab|$ to quantify the degree of entanglement of the time evolved state $\ket{\Phi(t)}$.
The concurrence, which completely determines the degree of entanglement of pure binary states \cite{Horodecki:2009review,Amico:2008review}, vanishes for separable states and is unity for maximally entangled states. 
Fig. \ref{fig:evolution}  shows the evolution of concurrence for a pair of molecules separated by $R = 10\,R_0$ under the action of a strong off-resonant Gaussian pulse. The pulsewidth $\tau_{\rm p}$ is chosen equal to the dipole-dipole interaction time $t_{\rm dd}$, while the peak intensity $\Omega_0$ is varied. 
Figure \ref{fig:evolution} shows that molecular entanglement is created in the presence of the laser pulse and reaches an asymptotic constant value when the pulse is over. We find that the qualitative behaviour of the system evolution is independent of $R$, $\Omega_0$ and $\tau_{\rm p}$, but that the actual value of the asymptotic concurrence depends strongly on the choice of these parameters.

Fig. \ref{fig:distance dependence} shows how the asymptotic concurrence $C(\rho)$ depends on the intermolecular distance $R$, or equivalently on the interaction time $t_{\rm dd}$, for fixed pulse parameters $\tau_{\rm p}=10^6\,t_{\rm R}$ and $\Omega_0=270$. We have chosen the pulse parameters here to ensure that two molecules separated by $R=100\,R_0$ ($R<R_{\rm e}$) become maximally entangled ($C(\rho)=1$). For smaller distances $R\leq 100\,R_0$, the asymptotic concurrence has an oscillatory dependence on $R$. 
For such distances the pulsewidth $\tau_{\rm p}$ is longer than the corresponding interaction time $t_{\rm dd}$. The system undergoes Rabi-type oscillations between the states $\ket{g_1g_2}$ and $\ket{e_1e_2}$ while the pulse is on. The oscillation stops when the pulse is over, giving the asymptotic concurrence shown in Fig. \ref{fig:distance dependence}. 
For larger distances $R>100 \,R_0$, the concurrence decays monotonically with $R$, and eventually for $R\gg R_{\rm e}$ there is no entanglement. In this case the pulsewidth $\tau_{\rm p}$ is smaller than $t_{\rm dd}$, and the state population does not have time to undergo a Rabi cycle.
Our calculations show that the behaviour of the asymptotic concurrence in Fig. \ref{fig:distance dependence} is independent of the choice of pulse parameters $\Omega_0$ and $\tau_{\rm p}$.
The fast decay of the entanglement with distance is particularly useful for an array of molecules. By choosing the laser pulse parameters $\Omega_0$ and $\tau_{\rm p}$ appropriately, it is possible to prepare highly entangled states between nearest neighbours only. 

\begin{figure}[t]
\begin{center}
\includegraphics[width=0.70\textwidth]{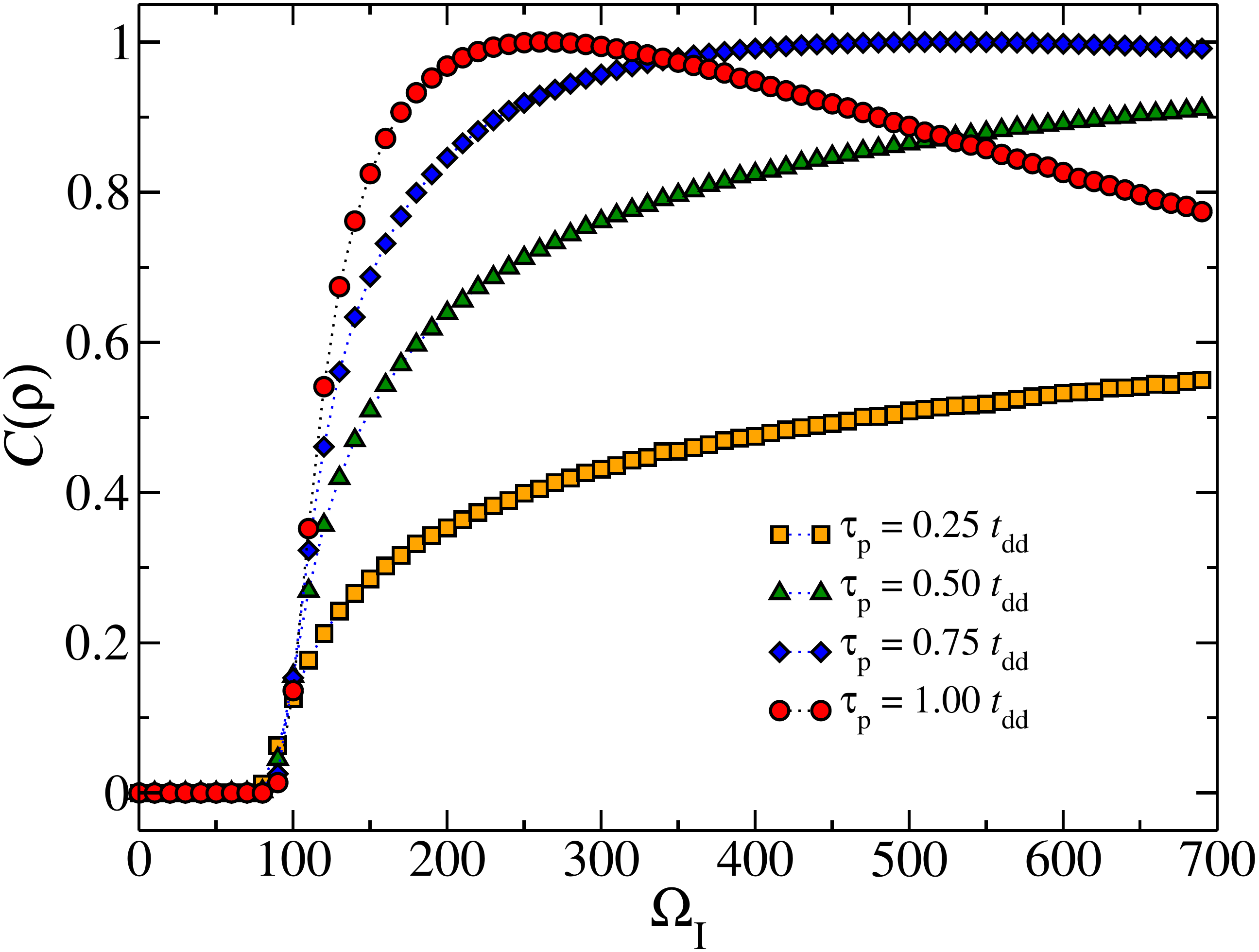}
\caption{Asymptotic concurrence $C(\rho)$ as a function of the peak intensity parameter $\Omega_{\rm I}$, long after the action of a Gaussian off-resonant laser pulse. 
The intermolecular distance is $R = 100 \,R_0$. Data is shown for different pulsewidths (FWHM): $\tau_{\rm p} = t_{\rm dd}$ (circles), $\tau_{\rm p} = 3t_{\rm dd}/4$ (diamonds), $\tau_{\rm p} = t_{\rm dd}/2$ (triangles), and $\tau_{\rm p}=t_{\rm dd}/4$ (squares). $t_{\rm dd} = 10^6 t_{\rm R}$ is the dipole-dipole interaction time and $t_{\rm R}=\hbar/B_{\rm e}$ is the rotational timescale.}
\label{fig:intensity dependence}
\end{center}
\end{figure}

The dependence of the asymptotic concurrence $C(\rho)$ on the laser pulse peak intensity $\Omega_0$ is shown in Fig. \ref{fig:intensity dependence}. Data 
are shown for a fixed distance $R =100\,R_0$ and for different values of the pulsewidth $\tau_{\rm p}$. For all values of $\tau_{\rm p}$, the concurrence is negligibly small below an intensity threshold, here $\Omega_0\approx 70$,
whose value depends on the intermolecular distance $R$. Independently of the pulsewidth, the asymptotic concurrence increases with the intensity above this threshold until it reaches the maximum value ($C(\rho) = 1$). For a given distance $R$, the maximum concurrence is achieved at smaller peak intensities $\Omega_0$ when the pulsewidth is equal to the dipole-dipole interaction time $t_{\rm dd}$. After reaching the maximum value, the concurrence decreases with intensity as the population of the doubly excited state $\ket{e_1e_2}$ exceeds $|b(t)|^2=1/2$ in Eq. (\ref{eq:final state}). In the strong field limit $\Omega_0\rightarrow\infty$, when $R$ and $\tau_{\rm p}=t_{\rm dd}$ are held constant, 
the population is completely transferred from $\ket{g_1g_2}$ to $\ket{e_1e_2}$, with no net entanglement creation.

The presence of an intensity threshold for the creation of molecular entanglement in Fig. \ref{fig:intensity dependence} can be related to the notion of entanglement radius $R_{\rm e}$ described earlier. For molecules within this radius, the mixing of the ground state $\ket{g_1g_2}$ with the two-excitation state $\ket{e_1e_2}$ is energetically favourable since the energy ratio $J_{12}/2\varepsilon_{\rm e}=\gamma(1-3\cos^2\Theta) (R_{\rm e}/R)^3$ exceeds unity. When this energy ratio is less than unity, the state mixing is suppressed and the concurrence becomes negligible. For a given distance $R$ and pulsewidth $\tau_{\rm p}$, the intensity threshold thus occurs at values of $\Omega_0$ for which $R_{\rm e}/R\sim 1$. In Fig. \ref{fig:intensity dependence}, $R_{\rm e}\approx 100\,R_0$ for $\Omega_0 = 130$. 

\subsection{Example: alkali-metal dimers in optical lattices}

\begin{table}[t]
\begin{center}
\begin{tabular}{cccccccc}\hline\hline
	& \; &	$d$ 	&$\Delta\alpha_{\rm V}$ &	$B_{\rm e}$ 		& $I_0$  		&$R_0$		& $t_{\rm R}$\\
	 & &	(D)	&	($a_0^3$)	&	(cm$^{-1}$)	&($10^{8}$ W/cm$^2$)	&(nm)		& (ps)	 	\\\hline
RbCs	 & &	1.238	&	441		&	0.0290		& 0.4			& 6.4		& 1.15		\\
KRb	 & &	0.615	&	360		&	0.0386		& 0.7			& 3.7		& 0.86		\\
LiCs	 & &	5.529	&	327		&	0.1940		& 3.8			& 9.3		& 0.17		\\
LiRb	 & &	4.168	&	280		&	0.2220		& 5.0			& 7.3		& 0.15		\\\hline\hline 
\end{tabular} 
\caption{Molecular parameters for selected polar alkali-metal dimers: $I_0$ is the laser intensity corresponding to $\Omega_{\rm I} \equiv \left({4\pi}/{c}\right){I_0\Delta\alpha_{\rm V}}/{2B_{\rm e}}=1$. $R_0=(d^2/B_{\rm e})^{1/3}$ is the characteristic length of the dipole-dipole interaction and $t_{\rm R}=\hbar/B_{\rm e}$ is the timescale of the rotational motion. Values of the polarizability anisotropy $\Delta\alpha_{\rm V}$, dipole moment $d$ and rotational constant $B_{\rm e}$ are taken from Ref. \cite{deiglmayr:2008-alignment}.}
\end{center}
\label{tab:intensities}
\end{table}

Table \ref{tab:intensities} lists the laser intensity $I_0$ of a traveling wave corresponding to a light-matter interaction parameter $\Omega_{\rm I}=1$ for selected polar alkali-metal dimers that have been optically trapped at ultracold temperatures \cite{Carr:2009,Chotia:2012}. Predicted values for the polarizability anisotropy $\Delta\alpha_{\rm V}$ and rotational constants for the rovibrational ground state are taken from Ref. \cite{deiglmayr:2008-alignment}. For alkali-metal dimers, $I_0$ is on the order of $10^7-10^8$ W/cm$^2$. 
This is well within the realm of feasibility, since continuous-wave laser beams with frequencies in the mid-infrared region ($\lambda\sim 1\,\mu$m) can have intensities on the order of $10^8$ W/cm$^2$ when focused to micrometer size regions \cite{Sugiyama:2007, Rungsimanon:2010}, while intensities higher than $10^{10}$ W/cm$^2$ can be achieved using pulsed lasers. Strong laser pulses are routinely used in molecular alignment experiments, with pulse durations varying from less than a femtosecond to hundreds of nanoseconds \cite{Sakai:1999,Seideman:2005}. 
 
We now consider the interaction of pairs of polar molecules with a strong off-resonant pulse when the molecules are trapped in individual sites of an optical lattice. Typical experimental lattice site separations are in the range $a_L=400 - 1000$ nm \cite{Bloch:2005,Danzl:2009}. For most alkali-metal dimers in Table \ref{tab:intensities}, these distances correspond to $R \sim 10^2 \,R_0$. The results in Figs. \ref{fig:distance dependence} and \ref{fig:intensity dependence} therefore show that highly-entangled states of molecules in different lattice sites can be prepared using a single laser pulse. For example, two LiRb molecules separated by $a_L=730$ nm can be prepared in a maximally entangled state by using a single Gaussian pulse with peak intensity $I = 1.35\times 10^{11}$ W/cm$^2$ and pulsewidth $\tau_{\rm p}=t_{\rm dd} = 150$ ns. These laser parameters can be achieved using current technology \cite{Sakai:1999}. 
It is therefore possible to generate highly entangled states in currently available optical lattice realizations by choosing the appropriate combination of parameters $\Omega_0$ and $\tau_{\rm p}$, regardless of the molecular species.

\section{Detection of molecular entanglement in optical traps}
\label{sec:entanglement quantification}

In this section we discuss how the alignment-mediated entanglement created between polar molecules in different sites of an optically trapped molecular array may be observed experimentally.  We first show that the pairwise entanglement created in an ensemble of molecules as described in Sec.~\ref{sec:entanglement generation} gives rise to coherent oscillations in the microwave absorption line shape.  Thus the global entanglement of the ensemble may already be detected by measurement of the linear spectral response as a function of frequency. We then outline how the time dependence of an initially entangled state generated by a strong laser pulse that subsequently evolves under the free rotational Hamiltonian may be tracked using correlations between local orientation measurements and a Bell inequality analysis \cite{Milman:2007, Milman:2009}.  For pairwise entanglement of a pure state, this allows a direct measurement of the concurrence measure of entanglement for the 
initially entangled state. 
This second entanglement detection scheme requires either single site addressing resolution in an optical lattice or individual trapping in separate dipole traps. 
Such addressability is now possible for trapped atoms \cite{Wilk:2010,Isenhower:2010,Weitenberg:2011} and is a subject of much experimental effort for trapped molecules. In contrast, the first approach is more amenable to current technology because it requires only global and not individual addressing. 

To show how these two detection schemes work, we shall consider explicitly an ensemble of molecules trapped in individual sites of a double-well optical lattice.
Such lattices can be prepared by superimposing standing waves with different periodicity \cite{Sebby:2006,Anderlini:2006,Sebby:2007,Lee:2007,Folling:2007}. When the distance between two neighbouring double wells is a few times longer than the separation
between the double-well minima, the alignment-mediated entanglement operation described in Sec. \ref{sec:entanglement generation} can be designed such that only molecules within a single double-well become entangled. 
Separability between neighboring pairs is ensured by increasing the distance between adjacent double wells.
We consider identical independent molecular pairs here for simplicity. In practice, inhomogeneities in the entanglement preparation step would lead to a distribution of
concurrence values throughout the array. 

In the remainder of this section we discuss the detection of entangled pairs initially prepared at time $t = 0$ by a strong laser pulse in the pure state $\ket{\Phi_0} = a_0\ket{g_1g_2}+b_0\ket{e_1e_2}$ and show how we may measure the value of the initial concurrence, $C(\rho_0)=2|a_0b_0|$.  
For times $t>0$, each molecule of the pair evolves under the free rotational Hamiltonian $\hat H_R$ (Section~\ref{sec:ac fields}). The state component $\ket{e_1 e_2}$ therefore acquires a relative dynamical phase which may modify time-dependent observables but does not change the concurrence. Our analysis will show that we can effectively extract the initial state concurrence $C(\rho_0)$ from both the linear absorption spectrum and orientational Bell inequality measurements.

\subsection{Global entanglement measure in optical lattices}
\label{sec:global}

It is well known that the macroscopic response of an ensemble of particles to an external field is affected by the presence of entanglement in the system \cite{Amico:2008review}. In particular, thermodynamic properties such the heat capacity and magnetic susceptibility have been established as entanglement witnesses for spin chains \cite{Amico:2008review,Vedral:2008}. 
In this section we will identify the signatures of entanglement on the AC dielectric susceptibility of a gas sample of $\mathcal{N}$ identical molecules.  For simplicity we consider an ensemble of identical entangled pairs but the results can readily be generalized to many-particle entangled states. 

In the absence of DC or near resonant AC electric fields, an ensemble of rotating polar molecules is unpolarized. An applied electric field $\mathbf{E}(t)$ creates a polarization $\mathbf{P}(t)$.  To lowest order in the field, this polarization is given by 
\begin{equation}
 \frac{\mathbf{P}(t)}{\mathcal{N}} = \frac{i}{\hbar}\int_{-\infty}^t dt'\left\{ \langle \mathbf{d}(t')\mathbf{d}(t)\rangle_0 - \langle \mathbf{d}(t)\mathbf{d}(t')\rangle_0\right\}\cdot\mathbf{E}(t'),
 \label{eq:Kubo}
\end{equation}
where $\langle  {\cdots} \rangle_0$ denotes an expectation value with respect to the state of the ensemble in the absence of the external field.  
Typically the system is in a thermal state $\hat \rho = \mathcal{Z}^{-1}(\beta)\rme^{-\beta \hat H_0}$, where $\hat H_0$ is the field-free Hamiltonian, $\mathcal{Z}(\beta)=\text{Tr}\{\rme^{-\beta \hat H_0}\}$ is the partition function and $\beta^{-1} = k_BT$. For equilibrium states the autocorrelation function $\langle \hat A(t)\hat B(t')\rangle_0$ depends only on the time difference $\tau = t-t'$.  As noted above, for analysis of the entanglement after the strong laser pulse is switched off, the Hamiltonian $\hat H_0$ is given by the two-molecule Hamiltonian $\mathcal{H}$ in Eq. (\ref{eq:H second-quantized}) with $\Omega_I = 0$.

Given the polarization, Eq. (\ref{eq:Kubo}), the microwave susceptibility for a thermal ensemble can be written as \cite{Mukamel-book}
\begin{equation}
 \chi(\omega) = -\mathcal{N}P_0(\beta)\left(\frac{d^2}{3\hbar}\right)\frac{1}{\omega-\omega_{eg}+i\gamma_e},
 \label{eq:chi MW thermal}
\end{equation}
where $P_0(\beta)\leq 1$ is the thermal population of the rotational ground state $\ket{0,0}$, and $\gamma_e$ is decay rate of the rotational excited state $\ket{1,0}$. The absorption spectrum is given by 
\begin{equation}
A(\omega) = \mathcal{N} \frac{ P_0(\beta)(d^2/3) \Gamma_e}{\left[(\hbar\omega - 2B_{\rm e})^2+\Gamma_e^2\right]},
\label{eq:absorption_thermal}
\end{equation}
where $A(\omega)\equiv {\rm Im}\{\chi(\omega)\}$ and $\Gamma_e = \hbar\gamma_e$ is the transition linewidth. 
%
%

Let us now consider the microwave susceptibility for an ensemble of entangled pairs initially prepared in the pure state $\ket{\Phi_0} = a_0\ket{g_1g_2}+b_0\ket{e_1e_2}$.  Unlike the thermal case, the corresponding density matrix $\rho_0 = \ket{\Phi_0}\bra{\Phi_0}$ describes a non-stationary state, with coherences that evolve according to $\hat H_0$ (in the absence of external perturbations). In this case the response of the system to the field $\mathbf{E}(t)$ is given by Eq. (\ref{eq:Kubo}) as for the thermal case, but the autocorrelation function $\bra{\Phi_0}\mathbf{d}(t)\mathbf{d}(t')\ket{\Phi_0}$ now depends on the absolute values of the time arguments $t$ and $t'$, where these are defined with respect to a common initial time.

The eigenstates of the coupled pairs in the limit $J_{12}/2\varepsilon_{\rm e}\ll 1 $ are $\ket{\Phi_1} = \ket{g_1g_2}$ with energy $E_1=0$, $\ket{\Psi_A} = 2^{-1/2}\left[\ket{g_1e_2} - \ket{e_1g_2}\right]$ with energy $E_A = \varepsilon_{e}- J_{12}$, $\ket{\Psi_S} = 2^{-1/2}\left[\ket{g_1e_2} + \ket{e_1g_2}\right]$ with energy $E_S=\varepsilon_{e}+ J_{12}$, and $\ket{\Phi_4} = \ket{e_1e_2}$ with energy $E_4=2\varepsilon_{\rm e}$ (see Eq. (\ref{eq:adiabatic states})). The energetic ordering of the states $\ket{\Psi_A}$ and $\ket{\Psi_B}$ depends on the sign of $J_{12}$. Using the non-stationary state $\Phi_0$ in the Kubo formula of Eq. (\ref{eq:Kubo}), the microwave absorption spectra at frequencies $\omega\approx \omega_{S1} \equiv(E_S-E_1)/\hbar$ can be written as
\begin{eqnarray}
A(\omega)&=&\mathcal{N}_{\rm P}\left(\frac{2d^2}{3\hbar}\right)\left[|a_0|^2\frac{\gamma_S}{(\omega_{S1}-\omega)^2+\gamma_S^2} \right.\nonumber\\
 &&\left. +|a_0b_0|\frac{\mathcal{F}_\omega(t)}{(\omega_{S1}-\omega)^2+\gamma_S^2}\right]  ,
 \label{eq:absorption dimer}
\end{eqnarray}
where $\mathcal{N}_{\rm P} = \mathcal{N}/2$ is the number of pairs, $\gamma_S$ is the decay rate of the state $\Psi_S$. In the derivation of Eq. (\ref{eq:absorption dimer})we have used the transition dipole moments 
$\bra{\Psi_S}\mathbf{d}\ket{\Phi_1} = \sqrt{2}\bra{e}\mathbf{d}\ket{g}=\bra{\Phi_4}\mathbf{d}\ket{\Psi_S}$, and $\bra{\Psi_A}\mathbf{d}\ket{\Phi_1}=0=\bra{\Phi_4}\mathbf{d}\ket{\Psi_A}$.
The function  $\mathcal{F}_\omega(t)$ contains the time dependence from the evolution of the entangled state under $\hat H_0$ and can be written as
\begin{equation}
\mathcal{F}_\omega(t) = \rme^{-\gamma_{41}t}\left[(\omega_{S1}-\omega)\sin\phi_{41}(t)+\gamma_S \cos\phi_{41}(t)\right],
\label{eq:dynamical lineshape}
\end{equation}
where $\phi_{41}(t) = \omega_{41}t-\theta_{ba}$ is the free phase evolution of the two-molecule coherence, $\theta_{ba}$ is the relative phase of the two components of the initial state, defined by $a_0^*b_0 = |a_0b_0|\rme^{i\theta_{ba}}$, and $\gamma_{41}$ is a decoherence rate introduced to account for dephasing channels. 

The amplitude of the time-dependent lineshape depends on the magnitude of the two-molecule coherence $|a_0b_0|=C(\rho_0)/2$.
For a maximally entangled two-molecule state $\ket{\Phi_0}$ with relative phase $\theta_{ba} = 0$, the peak absorption (per molecule) at the resonance frequency $\omega = \omega_{S1}$ is 
\begin{equation}
 \frac{A(\omega_{S1})\Gamma_S}{\mathcal{N}} = \frac{d^2}{6}\left[1+\cos(2\omega_{eg}t)\right].
\end{equation}
%

The presence of dynamical peaks in the absorption or emission spectra is a general feature of wavepacket evolution that has been widely studied for single atoms and molecules \cite{Mukamel-book}. More recently, the coherent oscillation of spectral peaks in the {\it nonlinear} optical response of molecular aggregates has been associated with entanglement between molecular units \cite{Sarovar:2010,Ishizaki:2010}. Equation (\ref{eq:absorption dimer}) shows that it is possible to identify entanglement in an ensemble of dipolar molecular pairs by measuring the linear absorption spectra. The procedure would be as follows.  After preparing the system in an entangled state using a strong off-resonant laser pulse, a weak microwave field tuned near resonance with the lowest dipole-allowed transition would give an absorption spectrum whose line width shows damped oscillations at frequency $\omega_{41} = 4B_{\rm e}/\hbar$. The presence of oscillations serves as an entanglement witness. Eq.~(\ref{eq:dynamical lineshape}) 
shows that the amplitude of this linewidth oscillation 
is proportional to the concurrence $C(\rho_0)=2|a_0b_0|$ of the initially 
prepared state, while the decay of the oscillation depends on the decoherence rate $\gamma_{41}$. 
Measuring the amplitude of these oscillations can thus allow measurement of the pairwise entanglement between the dipolar molecules.

\subsection{Bell's inequality for orientation correlations}

Bell inequalities quantify the differences between quantum and classical correlations of measurements performed in different  bases on quantum systems and provide critical tests of the incompatibility of quantum mechanics with local realism.  Violation of a Bell inequality constitutes evidence of nonlocal quantum correlations such as entanglement between distant particles \cite{Laloe:2001}.  Not all entangled bipartite states violate the inequality, although all separable states do satisfy the inequality \cite{Terhal:2000,Werner:2001}. For the case of entangled molecules in the presence of DC electric fields, it was recently shown that violations of Bell inequalities can be established~\cite{Milman:2007, Milman:2009}. In the following we adapt and simplify the analysis in Ref.~\cite{Milman:2007} to analize the orientational entanglement of polar molecules trapped in an optical double well lattice and prepared in the pure state $ \ket{\Phi_0} = a_0\ket{g_1g_2}+b_0\ket{e_1e_2}$ by the action of a strong off-
resonant 
laser pulse. We assume that the subsequent evolution is determined as in Sec.~\ref{sec:global} by the field-free rigid rotor Hamiltonian $\mathcal H_{\rm R}$, i.e., we neglect the small perturbation due to the trapping potential.

The degree of orientation of a single molecule is given by the expectation value of the operator $\hat O = \cos\theta$ \cite{Stapelfeldt:2003,Seideman:2005}, where $\theta$ is the polar angle of the internuclear axis with respect to the quantization axis. The orientation operator in the two-level basis $\mathcal{S}_1=\{\ket{g}\equiv\ket{0,0},\ket{e}\equiv\ket{1,0}\}$ can be written as $ \hat O = \sigma_X/\sqrt{3} $, with eigenvalues $\lambda_\pm = \pm 1/\sqrt{3}$, corresponding to the molecule being oriented parallel (plus sign) or antiparallel (minus sign) to the direction of the quantization axis.  
For our proposed realization with molecules trapped in double well optical lattices, orientation measurements can be performed in a using laser-induced fluoresence \cite{Orr-Ewing:1994} with single-site resolution.

We consider the two-time orientation correlation function for a molecular pair $ E(t_1,t_2)=\langle\hat{{O}}_1(t_1)\otimes\hat{{O}}_2(t_2)\rangle$,
where $\hat{{O}}_i(t_i) = \hat U_i^{\dagger}(t_i)\hat O(0)\hat U_i(t_i)$ \cite{Milman:2007, Milman:2009,Lemeshko:2011}. The free evolution operator is given by $\hat U(t) = \rme^{-i\hat H_{\rm R}t/\hbar}$, where $\hat H_{\rm R}=B_{\rm e} \sigma_Z$ in the two-level basis. The orientation correlation vanishes for separable two-molecule states, but remains finite for entangled states. In particular for a pair of molecules initially in the  state $\ket{\Phi_0}=a_0\ket{g_1g_2}+b_0\ket{e_1e_2}$, the orientation correlation function is given by
\begin{equation}
 E(t_1,t_2) = \frac{1}{3}C(\rho_0)\cos\left(\omega_{eg}t_1+\omega_{eg}t_2+\theta_{ba}\right), 
\label{eq:correlation function}
\end{equation}
where $C(\rho_0)$ is the concurrence of the initial pure state $\rho_0=\ket{\Phi_0}\bra{\Phi_0}$,
$\theta_{ba}$ the relative phase between the state components (see above) and the rotational frequency is $\omega_{eg} = 2B_{\rm e}/\hbar$. The correlation function is invariant under particle exchange and symmetric around $t_1=t_2=\pi/2$ for the relative phase $\theta_{ba}=n\pi$, with $n$ an integer.

Bell measurements can be divided into three steps \cite{Laloe:2001}. First is the preparation of a pair of particles, typically spins, in a repeatable way. Second, an experimental setting is chosen independently for each particle. The setting for spins corresponds to the orientation of a Stern-Gerlach apparatus that measures the spin projections of particles A and B along the directions $\vec{a}$ and $\vec{b}$, respectively. Finally, the correlation $E(\vec{a},\vec{b})$ between the measurement outcomes for different sets of directions $(\vec{a},\vec{b})$ are collected. For quantum correlation the Bell's inequality in the Clauser-Horne-Shimony-Holt  form \cite{CHSH:1969,Horodecki:2009review}
\begin{equation}
 |E(\vec{a},\vec{b})+E(\vec{a},\vec{b}')+E(\vec{a}',\vec{b})-E(\vec{a}',\vec{b}')|\leq 2\lambda^2_{\rm max}
\label{eq:Bell inequality}
\end{equation}
is violated, where $\lambda_{\rm max}$ is the maximum value of the measurement outcome. The quantum mechanical spin projection operator is $\vec{a}\cdot\vec{\sigma}$, with $\vec{\sigma}=(\sigma_X,\sigma_Y,\sigma_Z)$. For spin-$1/2$ particles $\lambda_{\rm max}=1$. 

\begin{figure}[t]
\begin{center}
\includegraphics[width=0.70\textwidth]{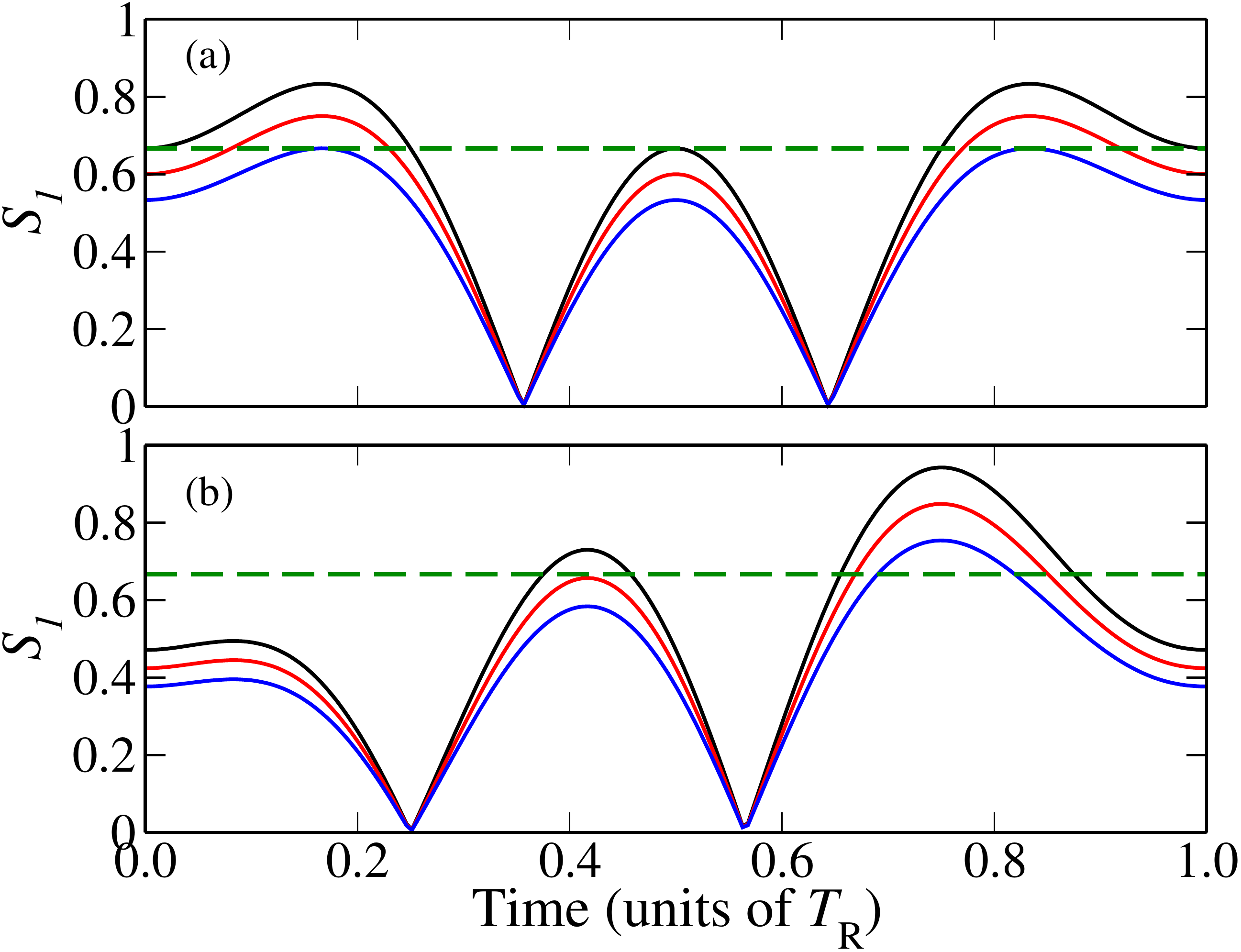}
\caption{Violation of Bell's inequality for molecular orientation correlations. The absolute value of $S_1(t)=  E(0,0)+E(0,t)+E(t,0)-E(t,t)$ is plotted as a function of time for several  states of the form $\ket{\Phi} = |a|\ket{g_1g_2}+|b|\rme^{i\theta_{ba}}\ket{e_1e_2}$. Each panel shows $|S_1|$ for three values of the concurrence: $C =1.0$ (black line), $C=0.9$ (red line), and $C=0.8$ (blue line). Panels (a) and (b) correspond to the relative phases $\theta_{ba}=0$ and $\theta_{ba}=\pi/4$, respectively. $E(t,t')$ is the two-time orientation correlation function. Time in is units of the rotational period $T_R = \pi\hbar/B_{\rm e}$.}
\label{fig7:Bell violation}
\end{center}
\end{figure}

There is a one-to-one correspondence between Bell measurements based on spin orientations $\vec{a}$ and $\vec{b}$ and a scheme based on the free rotational evolution of molecules.
In the two-state basis used here, the molecular orientation operator in the Heisenberg picture can be written as $ \hat O(\tau_a) = \frac{1}{\sqrt{3}}\rme^{i\hat{\sigma}_z\tau_a/2}\;\hat{\sigma}_X\;\rme^{-i\hat{\sigma}_z\tau_a/2}\equiv \vec{a}\cdot\vec{\sigma}$,
where we have defined the orientation vector $\vec{a} = (1/\sqrt{3})(\cos\tau_a,-\sin\tau_a,0)$, and $\tau_a=2B_{\rm e}t_a/\hbar$. 
The time evolution of the orientation operator $\hat O(\tau_a)$ thus corresponds to a clockwise rotation of the orientation direction $\vec{a}$ from the positive $X$ axis by an angle $ \tau_a$  in the $XY$ plane. Therefore, choosing the time $t_a$ when to perform a molecular orientation measurement is equivalent to choosing the orientation of the Stern-Gerlach apparatus for the case of spin-$1/2$ particles. The two-time orientation correlator in Eq. (\ref{eq:correlation function}) can thus be written as $E(t_a,t_b)=\langle\vec{a}\cdot\vec{\sigma}\otimes\vec{b}\cdot\vec{\sigma}\rangle$, which is the form of the correlation function for spin systems. Following the equivalence between spin 
orientation and rotational evolution, the magnitude of the quantity 
\begin{equation}
 S = E(t_a,t_b)+E(t_a,t_b')+E(t_a',t_b)-E(t_a',t_b').
\label{eq:rotational inequality}
\end{equation}
can then be used to test violations of Bell's inequality. For our purposes it is sufficient to set $t_a=t_b=0$ and $t_a' = t_b' = t$ in Eq. (\ref{eq:rotational inequality}) and evaluate the absolute value of $S_1(t) = E(0,0)+E(0,t)+E(t,0)-E(t,t)$ using Eq. (\ref{eq:correlation function}). In Fig. \ref{fig7:Bell violation} we plot $|S_1(t)|$ as a function of time for several parent states $\ket{\Phi_0}=a_0\ket{g_1g_2}+b_0\ket{e_1e_2}$ with different concurrences $C(\rho)$ and relative phases $\theta_{ba}$. The upper bound imposed by Bell's inequality over the $|S_1(t)|$ is $2\lambda_{\rm max}^2=2/3$. For the states shown in Fig. \ref{fig7:Bell violation}, this limit is violated over a wide range of times 
within a rotational period $T_{\rm R}=\pi t_{\rm R}$. The violation of the classical bound serves as an entanglement witness. Most importantly, the figure clearly shows that the degree of violation of Bell's inequality depends on the concurrence $C(\rho_0)$ of the entangled state. Therefore, once the signal is calibrated it should be possible to use the magnitude of $S_1(t)$ at a chosen time to quantify the molecular entanglement.

We close with some comments on experimental feasibility of these measurements. The preparation of entangled pairs can be done using the methods described in Sec. \ref{sec:entanglement generation}. An ensemble of identical pairs can be prepared to enhance the sensitivity of the correlation measurements. Performing orientation measurements in individual sites with laser-induced fluorescence \cite{Orr-Ewing:1994} is significantly less destructive than femtosecond photodissociation measurements.
Experimental violations of Bell's inequality have been established in a large number of experiments using photons \cite{Freedman:1972,Aspect:1981,Zeilinger:1998,Gisin:1998,Gisin:2001}, trapped atoms \cite{Rowe:2001}, superconducting junctions \cite{Ansmann:2009}, quantum dots \cite{Sun:2012},  and even elementary particles \cite{Apostolakis:1998}, but to the best of our knowledge it has not been established with molecules. Our analysis shows that it is possible with current technology to look for violations of Bell's inequality for molecules in long-wavelength optical lattices or in separate dipole traps. 

\section{Robustness of entanglement against motional decoherence}
\label{sec:decoherence}

Entanglement between distant molecules can be expected to decay in time due to relaxation and dephasing processes resulting from environmental perturbations. 
For entangled molecules in optical traps decoherence processes arise from their interaction with noisy external fields. 
Far-detuned optical traps, for example, are sensitive to laser intensity fluctuations and beam pointing noise, which can cause heating of the trapped atoms or molecules \cite{Savard:1997,Gehm:1998}. Trap noise affects the precision of atomic clocks \cite{Takamoto:2005,Ludlow:2006} and also the dynamics of strongly-correlated cold atomic ensembles \cite{Pichler:2012}. Additional sources of decoherence influence the dynamics of the system in the presence of static electric and magnetic fields \cite{Yu:2003}. 
In this Section we analyze the robustness of alignment-mediated entanglement of molecules trapped in optical lattices to fluctuations in the optical trapping laser fields. Our primary focus here is on motional decoherence in optical arrays, which is most sensitive to the effective lattice temperature. 

For an array of interacting polar molecules, the fluctuation of the dipole-dipole interaction energy $U_{\rm dd}(R)$ with the motion of the molecules in the trapping potential represents a source of decoherence for the collective rotational state dynamics. The vibrational motion of the molecules in an optical lattice potential can be represented by phonons interacting with the coherent rotational excitation transfer between molecules in different sites. Following Ref. \cite{Herrera:2011} we write the Hamiltonian for a one-dimensional molecular array in the absence of static electric fields as
\begin{eqnarray}
  \mathcal{H} &=&  \sum_i\epsilon_{eg} \creation{c}{i}\annihilation{c}{i} + \sum_{i,j} J_{ij} \creation{c}{i}\annihilation{c}{j} \nonumber\\
  &&+ \sum_k \hbar\omega_k \crea{a}{k}\anni{a}{k}+\sum_{i,j\neq i}\sum_k \lambda_{ij}^k \creation{c}{i}\annihilation{c}{j}\left(\annihilation{a}{k} + \creation{a}{k}\right),
\label{eq:lattice Hamiltonian}
\end{eqnarray}
where $\creation{a}{k}$ creates a phonon in the $k$-th normal mode with frequency $\omega_k$. The first and second terms determine the coherent state transfer between molecules in different sites, with site energy $\epsilon_{eg}=2B_{\rm e}$ and hopping amplitude $J_{ij}$ (evaluated at equilibrium distances). The third term describes the vibrational energy of the molecular center of mass in the trapping potential, which we assume harmonic as an approximation. In the absence of DC electric fields the phonon spectrum is dispersionless \cite{Herrera:2011}, i.e., $\omega_k = \omega_0$. The last term represents the interaction between the internal and external molecular degrees of freedom, characterized by the energy scale
\begin{equation}
\lambda_{ij}^{k}(\omega_0)=-3J_{12}\left[\frac{l_0(\omega_0)}{a_L}\right]f^k_{ij}\frac{(i-j)}{|i-j|^{5}}, 
\label{eq:lambda}
\end{equation}
where $\omega_0$ is the trapping frequency
of the optical lattice, $a_L$ is the lattice constant, $l_0 = \sqrt{\hbar/2m\omega_0}$ is the oscillator length, and $f^k_{ij}$ is a mode-coupling function that satisfies the relation $f_{ij}^k=-f_{ji}^k$. 

We have omitted terms of the form $(\creation{c}{i}\creation{c}{j} + \text{H.c})$ in Eq. (\ref{eq:lattice Hamiltonian}), since 
these only affect the dynamics of the system when $J_{12}/\epsilon_{eg}\sim 1$. As discussed in Section \ref{sec:entanglement generation}.1, this condition is satisfied only in the presence of a strong off-resonant pulse. However, the laser pulse width $\tau_p$ is orders of magnitude shorter than the timescale of the oscillation of molecules in the lattice potential ($\tau_{\rm p}\ll \omega_0^{-1}$). This separation of timescales allows us to neglect the coupling between internal and translational degrees of freedom under the action of a strong off-resonant laser pulse, even when $J/\epsilon_{eg}\sim 1$. After the pulse is over, the coupling to phonons can become important when the timescale for internal state evolution $h/J_{12}$ is comparable with $1/\omega_0$. Under this condition the molecular array evolves according to the Hamiltonian in Eq. (\ref{eq:lattice Hamiltonian}) over a timescale shorter than the molecular trapping lifetime $\tau_{\rm trap}\sim 1 $ s \cite{Chotia:2012}. 

The Hamiltonian in Eq. (\ref{eq:lattice Hamiltonian}) can be rewritten as $\mathcal{H} = \mathcal{H}_S+\mathcal{H}_B+\mathcal{H}_{SB}$ using the unitary transformation $\crea{c}{\mu} = \sum_i u_{i\mu}\crea{c}{i}$. The Hamiltonian $\mathcal{H}_S = \sum_\mu \varepsilon_\mu \crea{c}{\mu}\anni{c}{\mu}$ describes the collective rotational states in terms of excitonic states $\ket{\mu} = \crea{c}{\mu}\ket{g}$ with energy $\varepsilon_\mu$. The second term $\mathcal{H}_B=\hbar\omega_0\sum_k\crea{a}{k}\anni{a}{k}$ describes free lattice phonons, and the term 
\begin{equation}
\mathcal{H}_{SB} = \sum_{\mu\nu}\lambda_{\mu\nu}^k\crea{c}{\mu}\anni{c}{\nu}(\anni{a}{k}+\crea{a}{k}),
\label{eq:system-bath}
\end{equation}
describes the interaction of the excitonic system with the phonon environment. The interaction energy in the exciton basis is given by $\lambda^k_{\mu\nu} = \sum_{ij}u^*_{i\mu}u_{j\nu}\lambda_{ij}^k$. 
The internal state evolution of the excitonic system depends strongly on the characteristics of the phonon environment. For low phonon frequencies $\omega_0<J_{12}/h$ the interaction energy $\lambda^k$ can become the largest energy scale in the Hamiltonian, and non-Markovian effects in the evolution of the system density matrix $\rho(t)$ become important \cite{Breuer-Petruccione-book}. 
We assume here for simplicity that $\hbar\omega_0>J_{12}$, or more precisely $(l_0/a_L)^2(J_{12}/\hbar\omega_0)< 1$ \cite{Herrera:2012} so that we are in a weak coupling regime.  
Note that $\omega_0$ is determined by the trapping strength of the optical lattice and that both this and the dipolar interaction $J_{12}$ can be tuned in this system to a far greater extent than is possible for Hamiltonians describing excitonic energy transfer in molecular aggregates~\cite{Agranovich:2008}. In this weak coupling regime, the system evolution can then be described by a quantum master equation in the Born-Markov and secular approximations \cite{Breuer-Petruccione-book}\footnote{Note that the secular approximation does not allow for coherence transfer \cite{Engel:2007}} as $\dot \rho(t) = -(i/\hbar)\left[ \mathcal{H}_S,\rho(t)\right]+\mathcal{D}\left(\rho(t)\right)$. 

Let us consider the case of two interacting polar molecules coupled to a common phonon environment via the nonlocal term in Eq. (\ref{eq:system-bath}). The dissipative dynamics of the system density matrix $\rho(t)$ is determined by 
\begin{equation}
\mathcal{D}(\rho(t)) = \gamma_0\mathcal{P}_1^{(-)}\rho(t) \mathcal{P}_1^{(-)}-\frac{1}{2}\gamma_0\{\mathcal{P}_1^{(+)},\rho(t)\},
\label{eq:dissipator nonlocal} 
\end{equation}
where $\mathcal{P}_1^{(\pm)}=\ket{\Psi_S}\bra{\Psi_S}\pm\ket{\Psi_A}\bra{\Psi_A}$ are projection superoperators, $\gamma_0$ is the pure-dephasing rate, and $\{A,B\}$ denotes the anticommutator. The projection into the two-excitation eigenstate $\mathcal{P}_2=\ket{e_1e_2}\bra{e_1e_2}$ does not contribute in the absence of DC electric fields (see discussion in \ref{sec:dc fields}). The single-excitation eigenstates are $\ket{\Psi_S}=2^{-1/2}(\ket{e_1g_2}+\ket{g_1e_2})$ and $\ket{\Psi_A} = 2^{-1/2}(\ket{e_1g_2}-\ket{g_1e_2})$. Equation (\ref{eq:dissipator nonlocal}) shows that for a system prepared in the pure state $\ket{\Phi}=a\ket{g_1g_2}+b\ket{e_1e_2}$ we have $\mathcal D(\rho) = 0$. In other words, the 
two-molecule entangled states prepared using a strong laser pulse do not decohere due to the interaction with environmental phonons in the optical lattice, regardless of the strength of the coupling to the environment and the effective lattice temperature. This is a consequence of the nonlocal nature of the interaction with the phonon environment
and implies that under these conditions, the states $\ket{\Phi_{\pm}}=\left[\ket{g_1g_2}\pm\ket{e_1e_2}\right]$ provide a basis for a decoherence-free subspace in which all pairwise entangled states may be defined.

We can understand the effects of motional decoherence on the entangled triparticle and many-particle states by estimating the full phonon decoherence rates, given by $\gamma_{\mu\nu,\mu'\nu'}(\omega)$, with $\mu, \nu$ indexing the excitonic states. In Eq. (\ref{eq:dissipator nonlocal}) the pure dephasing rate is defined as $\gamma_0=\gamma_{AA,AA}(0)=\gamma_{SS,SS}(0)=-\gamma_{AA,SS}(0)=-\gamma_{SS,AA}$(0).
In the Born-Markov and secular approximations, dephasing and relaxation processes that lead to decoherence and entanglement decay occur at the rate $\gamma_{\mu\nu,\mu'\nu'}(\omega) = (1/\hbar^2)\int_{-\infty}^\infty d\tau\rme^{i\omega\tau}\langle \hat B_{\mu\nu}(\tau)\hat B_{\mu'\nu'}(0)\rangle$, where $\langle \hat B_{\mu\nu}(\tau)\hat B_{\mu'\nu'}(0)\rangle$ is the bath correlation function with $\hat B_{\mu\nu} = \sum_k\lambda_{\mu\nu}^k(\anni{a}{k}+\crea{a}{k})$. In \ref{sec:spectral density} we use a classical stochastic model to approximate the bath correlation function 
under the influence of random intensity fluctuations of the trapping laser. This procedure allows us to write the decoherence rates as
\begin{equation}
\gamma_{\mu\nu,\mu'\nu'}(\omega) =  \frac{1}{\hbar^2}\left[n(\omega)+1\right]\left[J^{\rm cl}_{\mu\nu,\mu'\nu'}(\omega) - J^{\rm cl}_{\mu\nu,\mu'\nu'}(-\omega)\right],
 \label{eq:transition rate}
\end{equation}
where $n(\omega) = (\rme^{\beta\hbar\omega}-1)^{-1}$ is the Bose distribution function and 
\begin{equation}
 J^{\rm cl}_{\mu\nu,\mu'\nu'}(\omega) = \sum_k\lambda_{\mu\nu}^k\lambda_{\mu'\nu'}^k\left(\frac{\omega}{\omega_k}\right)\frac{\beta}{(\omega - \omega_k)^2+\beta^2},
\end{equation}
is the semiclassical spectral density for optical lattice phonons. In \ref{sec:spectral density} we show that the broadening parameter can be written as $\beta = \kappa \omega_0^2$, where the factor $\kappa>0$ is proportional to the strength of the laser intensity noise. The trapping noise causes damping of the correlation function as $\langle B_{\mu\nu}(t)B_{\mu\nu}(0)\rangle\propto\rme^{-\beta|t|}\cos(\omega't)$, where $\omega' = \sqrt{\omega_0^2-\beta^2}$. The bath autocorrelation time $\tau_c$ is order $\beta^{-1}$. The condition for the Markov approximation to hold is thus $\beta^{-1}\ll h/J_{12}$. 

For fixed trapping parameters $\omega_0$, $a_L$ and $\beta$, this analysis shows that different molecular species can undergo very different open system dynamics, depending on the strength of the dipolar interaction between molecules in different sites.  For instance, let us consider LiCs ($d=5.5$ D) and KRb ($d = 0.6$ D) species as examples of molecules with high and low permanent dipole moments, respectively. For an optical lattice with $a_L = 1\,\mu$m and noise-induced damping rate $\beta = 100$ Hz, the open system dynamics would have Markovian behaviour for KRb molecules ($J_{12}/h = 10 $ Hz), but for LiCs molecules ($J_{12}/h = 1.4$ kHz) the system dynamics can be expected to be non-Markovian. 
A very attractive feature of this trapped dipolar molecule array is that the transition between Markovian and non-Markovian dynamics can be studied experimentally for any molecular species by manipulating the laser intensity noise in order to tune the parameter $\beta$ as in Ref. \cite{DErrico:2012}, or by changing the lattice spacing $a_L$ to manipulate $J_{12}$. 

In the regime where the Markov and secular approximations are valid, we can estimate the phonon-induced decoherence rate $\gamma(\omega_S)$ in Eq. (\ref{eq:transition rate}) (with state indices removed for simplicity) at the characteristic system frequency $\omega_{\rm S} = J_{12}/\hbar$. For a lattice temperature such that $\hbar\omega_S/k_{\rm b}T\ll 1$ the decoherence rate scales as $\gamma(\omega_S)\sim 4\pi^2(J_{12}/h)^2(l_0/a_L)^2 H(\omega_S)$, with $H(\omega) = (\omega/\omega_0)\beta/[(\omega-\omega_0)^2+\beta^2]$. For experimentally realizable parameters $\beta = 1$ kHz, $\omega_0 = 10$ kHz and $a_L = 500$ nm, the decoherence rate for KRb molecules ($\omega_S/2\pi=0.13$ kHz) is $\gamma(\omega_S)\sim 10^{-5}$ Hz, which is negligibly small compared with the typical loss rate of molecules from optical traps ($\gamma_{\rm trap}\sim 1$ Hz) due to incoherent Raman scattering of lattice photons. 
We conclude that the entangled states of polar molecules containing double excitations can be robust to phonon-induced decoherence in optical lattice settings for which the weak coupling condition $\hbar\omega_0/J_{12}\gg 1$ holds.

\section{Conclusion}
\label{sec:conclusions}

In this work we present a scheme to generate entanglement in arrays of optically trapped polar molecules. Starting from an array of molecules prepared in their rovibrational ground state, a single strong off-resonant laser pulse can be used to generate entanglement between molecules in different sites of the array. The strong laser field induces the alignment of molecules along its polarization direction during the pulse. 
For such laser alignment of polar molecules interacting via a dipole-dipole term, the energy ratio between the coupling and site energies $J_{12}/\varepsilon_{\rm e}$ can be larger than unity, allowing generation of two-particle wavefunctions of the form $\ket{\Phi} = a\ket{g_1g_2}+b\ket{e_1e_2}$ in the presence of the strong laser field.  For $|ab|\neq 0$, the laser alignment will thus induce entangled states, where the precise form of the resulting entangled state may be controlled by the duration and strength of the laser pulse.  The subsequent evolution after the laser pulse is completed adds a dynamical phase to the entangled state but does not change the concurrence measure of the extent of entanglement.  
The proposed generation scheme does not depend on the number of coupled molecules and also holds for a many-particle system. Here for simplicity we have considered explicitly only the two-particle case.

We emphasize that this alignment-mediated entanglement involving double excitation states is not possible with static electric fields.  The rotational structure of an aligned molecule is such that the transition energy $\varepsilon_{\rm e}$ between the lowest two rotational states $\ket{g}$ and $\ket{e}$ becomes comparable in magnitude with the dipole-dipole interaction energy $J_{ij} =\bra{g_ig_j}\hat V_{\rm dd}\ket{e_ie_j}$, for molecules separated by distances of up to several micrometers. At such 
large distances the ratio $J_{12}/\varepsilon_{\rm e}$ is negligibly small in the absence of DC electric fields and double-excitation transitions of the type $\ket{g_1g_2}\rightarrow\ket{e_1e_2}$ are energetically suppressed.

We have demonstrated explicitly that the degree of entanglement in a molecular pair can be manipulated by tuning experimental parameters such as the laser pulse intensity and duration, as well as the intermolecular distance. We presented two methods to detect and measure entanglement in optical traps after the strong laser pulse is applied. 
The first approach requires only global microwave addressing of the molecular array. Here we showed that the linear microwave response of an ensemble of entangled pairs contains a contribution to the absorption lineshape that is proportional to the amount of pairwise entanglement and that oscillates in time at a frequency of order $B_{\rm e}/h$, where $B_{\rm e}$ is the rotational constant. Measuring the absorption peak oscillations over this timescale would then allow the concurrence of the state to be determined. 
The second approach is based on measurements of molecular orientation correlations to establish violations of Bell's inequality. This method relies on the ability to optically address individual sites of a molecular array in order to perform laser-induced fluorescence measurements.
Finally, we also analyzed the robustness of the strong field alignment-mediated molecular entanglement in 
optical arrays with respect to motional decoherence induced by fluctuations in the trapping lasers.  

The results presented in this work for a molecular pair can readily be generalized to larger molecular arrays, as indicated in the text of the paper.
In this context, it is useful to recognize that the system Hamiltonian can be mapped into a quantum-Ising model with a tunable magnetic field, a model that has been widely used in the study of quantum phase transitions \cite{Amico:2008review}. Furthermore, the form of Ising Hamiltonian describing the system is 2-local, which supports universal quantum computation when combined with the ability to implement arbitrary single-particle unitary transformations \cite{Lloyd:1995}. Therefore, an array of optically-trapped polar molecules driven by strong off-resonant laser pulses provides both a test-bed for studies of quantum entanglement in many-body systems and a novel platform for the development of quantum technologies.

\section*{Acknowledgements}

We thank Roman Krems for helpful comments on the manuscript. FH and SK were supported by the NSF CCI center ``Quantum Information for Quantum Chemistry (QIQC)'', award number CHE-1037992. FH was also supported by NSERC Canada.

\appendix

\section{Molecules in combined off-resonant laser and DC electric fields}
\label{sec:dc fields}
In this appendix we describe the dipole-dipole interaction between polar molecules in combined presence of DC electric fields and strong off-resonant pulsed laser fields.  We discuss how the addition of a DC electric field affects the entanglement creation scheme described in Section \ref{sec:entanglement generation}.

\subsection*{Dipole-dipole interaction in combined fields}

Let us consider a polar molecule in its vibrational ground state, under the influence of a DC electric field and a CW far-detuned optical field. If the laser polarization is collinear with the direction of the DC electric field (space-fixed $Z$ axis), the dimensionless molecular Hamiltonian $\hat H = \hat H_{\rm R} + \hat H_{\rm DC} + \hat H_{\rm AC}$ can be written in analogy with Eq. (\ref{eq:dimless ac}) as 
\begin{equation}
 \hat H = \hat N^2 -\lambda\mathcal{D}^{(1)}_{0,0}-\frac{2}{3}\Omega_{\rm I}\mathcal{D}^{(2)}_{0,0}, 
\label{app:dimless ac/dc}
\end{equation}
where $\lambda = dE_Z/B_{\rm e}$ parametrizes the strength of the DC electric field. $E_Z$ is the magnitude of the DC electric field and $d$ is the permanent dipole moment of the molecule. The rotational structure for $E_Z=0$ and large laser intensities $\Omega_{\rm I}$ consists of harmonically spaced tunneling doublets separated by an energy proportional to $\Omega_{\rm I}$ as shown in Fig. \ref{fig:ac energies} of the main text. Each doublet is composed of states with opposite parity whose energy splitting decreases exponentially with $\Omega_{\rm I}$. Due to this near degeneracy, a very weak DC electric field strongly couples the field-dressed doublet states, splitting their energy levels linearly with $\lambda$ \cite{Friedrich:1999}. The two lowest doublet states $\ket{g}$ and $\ket{e}$ for $\lambda\ll 1$ correlate adiabatically with $\ket{g} \approx \sqrt{a}\ket{0,0}+\sqrt{b}\ket{1,0}$ and $\ket{e} \approx \sqrt{b}\ket{0,0} - \sqrt{a}\ket{1,0}$ as $\Omega_{\rm I}\rightarrow 0$, with 
$a\gg b$ and $\ket{NM_N}$ is 
an eigenstate of $\hat H_{\rm R}$.

In the absence of DC electric fields the dipole-dipole interaction operator $\hat V_{\rm dd}$ has only one non-zero matrix element $J_{ij} = \bra{e_ig_j}\hat V_{\rm dd}\ket{g_ie_j}=\bra{e_ie_j}\hat V_{\rm dd}\ket{g_ig_j}$, defined in Eq. (\ref{eq:exchange coupling}). In the presence of DC electric fields the parity of the rotational states is broken and the following matrix elements become finite: $V_{ij}^{gg} = \bra{g_ig_j}\hat V_{\rm dd}\ket{g_ig_j}$, $V_{ij}^{ee} = \bra{e_ie_j}\hat V_{\rm dd}\ket{e_ie_j}$, and $V_{ij}^{eg} = \bra{e_ig_j}\hat V_{\rm dd}\ket{e_ig_j}$. The dipolar energies $\left\{J_{ij},V_{ij}^{gg},V_{ij}^{ee},V_{ij}^{eg}\right\}$ determine the dynamics of interacting polar molecules in the regime where the energy $\Delta \epsilon_{eg}$ for the transition $\ket{g}\rightarrow\ket{e}$ is much larger than the dipole-dipole energy $U_{\rm dd}=d^2/R^3$, where $R$ is the intermolecular distance. In the regime $\Delta\epsilon_{eg}\sim U_{\rm dd}$ two additional dipole-dipole transitions become 
important: $A_{ij} = \bra{e_ig_j}\hat V_{\rm dd}\ket{g_ig_j}$ and $B_{ij} = \bra{e_ig_j}\hat V_{\rm dd}\ket{e_ie_j}$. These matrix elements couple the single excitation manifold with the ground and doubly excited states, and vanish in the absence of DC electric fields.

\begin{figure}[t]
\begin{center}
\includegraphics[width=0.80\textwidth]{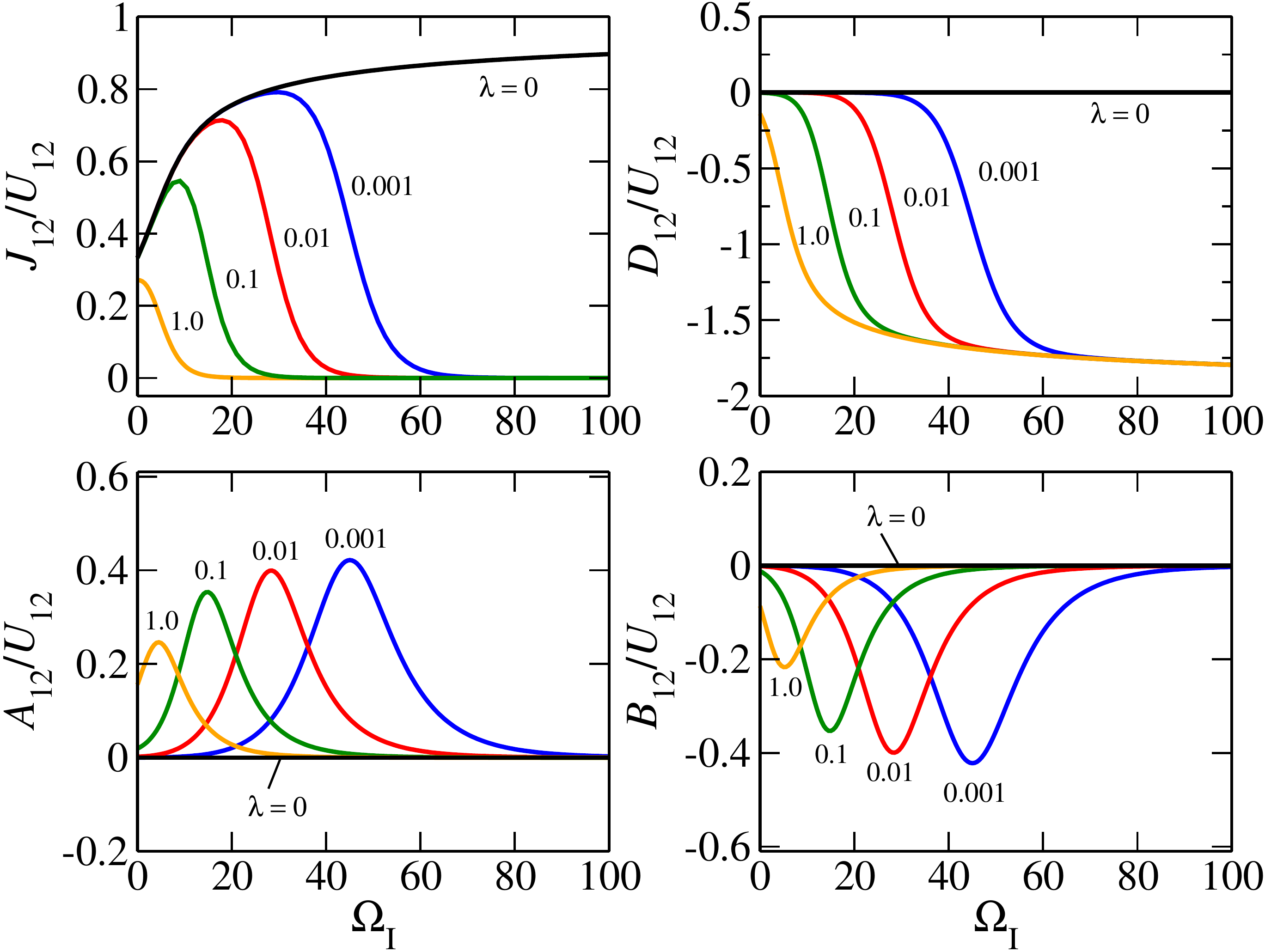}
\end{center}
\caption{ Dipole-dipole interaction energies $J_{12}$, $D_{12} \equiv V_{12}^{eg}-V_{12}^{gg}$, $A_{12}$ and $B_{12}$ as a function of the intensity parameter $\Omega_{\rm I}$. Curves are labeled according to the DC electric field strength $\lambda=dE_Z/B_{\rm e}$. The DC and AC electric fields are collinear. Energy is in units of $U_{\rm dd}=d^2/R^3$ and the intermolecular axis is taken perpendicular to the orientation of the fields.}
\label{fig:dipole energies}
\end{figure}

In analogy with the definition of $J_{ij}$ in Eq (\ref{eq:exchange coupling}) we can write the dipole-dipole energies in units of $U_{\rm dd}(1-\cos^2\Theta)$ as $V^{gg}_{ij} = \mu_g^2$, $V^{eg}_{ij} = \mu_{e}\mu_g$, $V^{ee}_{ij} = \mu_{e}^2$, $A_{ij} = \mu_{eg}\mu_g$, and $B_{ij} = \mu_{eg}\mu_e$, where $\mu_{eg} = d^{-1}\bra{e}\hat d_0\ket{g}$ is the dimensionless transition dipole, $\mu_e=d^{-1}\bra{e}\hat d_0\ket{e}$ is the dimensionless dipole moment of the excited state and $\mu_g=d^{-1}\bra{g}\hat d_0\ket{g}$ is the dipole moment of the ground state. For the choice of rotational states used here we have $\mu_{eg}>0$, $\mu_g>0$ and $\mu_e<0$, which give $A_{ij} = -B_{ij}>0$.
It is convenient to define the differential dipolar shift $D_{ij} = V^{eg}_{ij}-V^{gg}_{ij}=\mu_g(\mu_e-\mu_g)<0$ to describe the single-excitation dynamics \cite{Herrera:2011}. We evaluate the dipole-dipole matrix elements using the eigenvectors of the single-molecule Hamiltonian in Eq. (\ref{app:dimless ac/dc}). In Fig. \ref{fig:dipole energies} we show the dependence of the dipole-dipole energies $J_{ij}$, $D_{ij}$, $A_{ij}$ and $B_{ij}$ on the laser intensity parameter $\Omega_{\rm I}$ and the DC field strength parameter $\lambda$. The figure shows that the exchange interaction energy $J_{ij}$ tends to zero at high intensities $\Omega_{\rm I}\gg 10$ in the presence of a perturbatively small DC electric field $\lambda\ll 1$. The energies $A_{ij}$ and $B_{ij}$ also vanish at high intensities. Only the diagonal dipolar shifts $V_{ij}^{eg}$, $V^{ee}_{ij}$ and $V_{ij}^{gg}$ are finite in the high intensity regime for any non-zero DC field strength.

\begin{figure}[t]
\begin{center}
\includegraphics[width=0.70\textwidth]{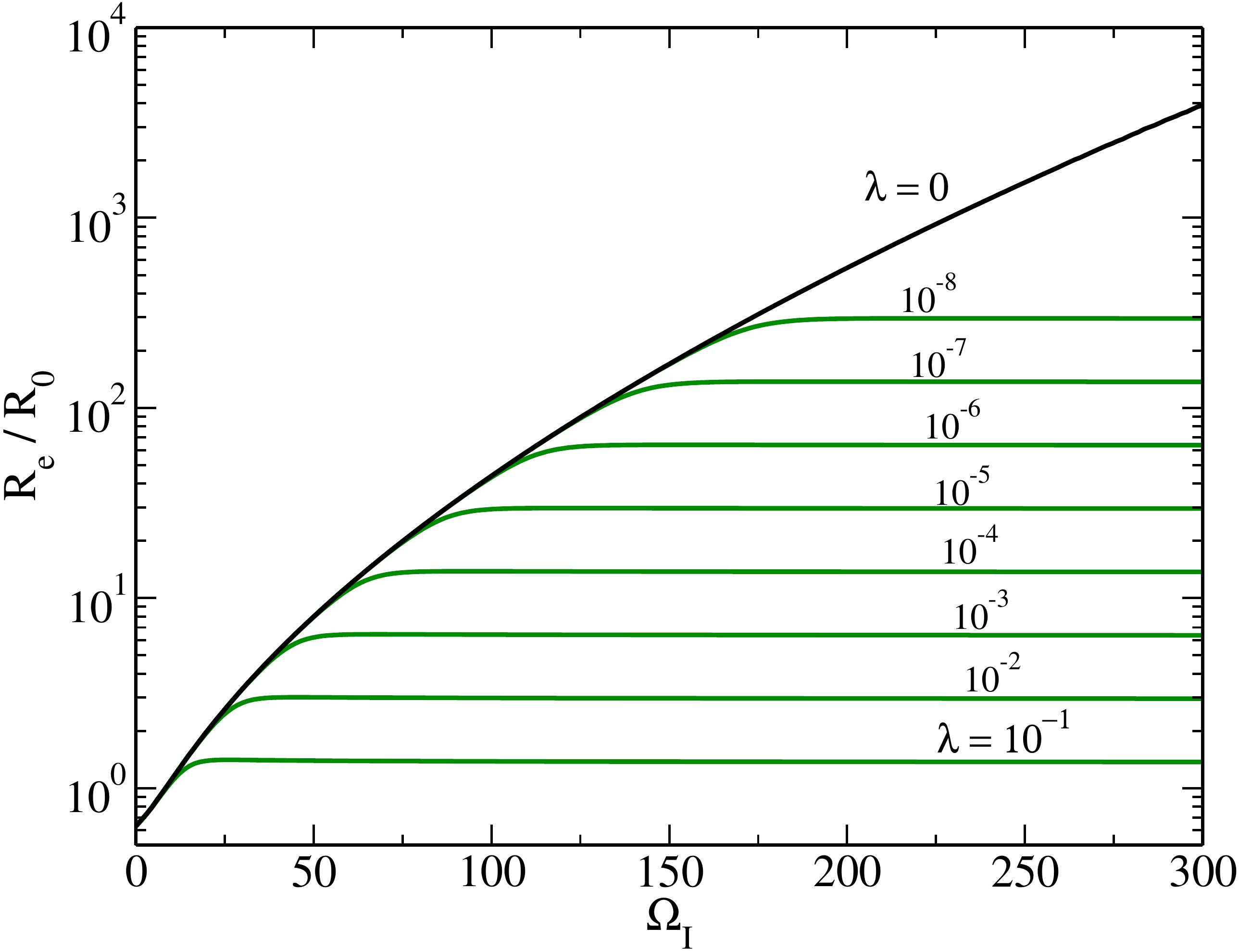}
\caption{Entanglement radius $R_{\rm e}$ in units of $R_0$ (log scale), as a function of the laser intensity parameter $\Omega_{\rm I}$. Curves are labeled according to the electric field strength $\lambda = dE_Z/B_{\rm e}$. $R_0\equiv(d^2/B_{\rm e})^{1/3}$ is a characteristic dipolar radius.}
\label{fig:entanglement length}
\end{center}
\end{figure}

\subsection*{Disadvantages for dynamical entanglement creation}

The presence of a DC electric field modifies the state evolution under the action of a strong off-resonant laser pulse in two ways. First, a static electric field strongly mixes the quasi-degenerate doublet states at high laser intensities (Fig. \ref{fig:ac energies}), resulting in a linear DC Stark shift that increases the energy splitting $\varepsilon_{\rm e}$. The Stark splitting significantly modifies the entanglement radius $R_{\rm e} = (d^2/2\varepsilon_{\rm e})^{1/3}$, as shown in Fig. \ref{fig:entanglement length}. The value of $R_{\rm e}$ increases exponentially with the laser intensity parameter $\Omega_{\rm I}$ in the absence of DC electric fields, but has an upper bound in combined fields. The bound depends on the DC field strength $\lambda = dE_Z/B_{\rm e}$, which determines the splitting of the states $\ket{g}$ and $\ket{e}$. For larger values of $\lambda$, the intermolecular distance at which the dipole-dipole interaction between molecules becomes comparable with the Stark splitting 
becomes smaller. For the molecular species used in Table \ref{tab:intensities}, $\lambda\sim 1$ corresponds to $E_Z\sim 1 $ kV/cm. For such large field strengths, $R_{\rm e}\approx R_0\sim 1$ nm for most alkali-metal dimers. Therefore, molecules in optical lattices with site separation $R\sim 10^2$ nm cannot be entangled using strong off-resonant fields when DC electric fields $E_Z\sim 1$ kV/cm are present. Figure \ref{fig:entanglement length} however shows that in the presence of stray fields $E_Z\leq 1$ mV/cm ($\lambda\leq 10^{-6}$), alignment-mediated entanglement of alkali-metal dimers in optical lattices is still possible. 

Second, breaking the parity symmetry of the rotational states results in additional contributions to the dipole-dipole interaction such as already discussed. The matrix elements $A_{ij}$ and $B_{ij}$ mix the subspaces $\mathcal{S}_1 = \left\{\ket{g_1e_2}, \ket{e_1g_2}\right\}$ and $\mathcal{S}_2 = \left\{\ket{g_1g_2}, \ket{e_1e_2}\right\}$, the two-molecule state for the initial condition $\ket{\Phi(0)} = \ket{g_1g_2}$ is given by $\ket{\Phi(t)} = a(t)\ket{g_1g_2}+b(t)\ket{e_1g_2}+c(t)\ket{g_1e_2}+d(t)\ket{e_1e_2}$, with $|ad|\neq 0$ and $|bc|\neq 0$. Therefore, for intermolecular distances $R\leq R_{\rm e}$ the two-molecule state evolution in combined DC and off-resonant fields no longer follows the simple two-state dynamics described in Section \ref{sec:entanglement generation}.

Static electric fields also affect the dynamics of the entangled states after the laser pulse is over. Local system-environment coupling occurs in the presence of a static electric field \cite{Herrera:2011}. The local interaction of a pair of molecules with the phonon environment is described by $\hat H_{\rm int} = \kappa (\creation{c}{1}\annihilation{c}{1}+\creation{c}{2}\annihilation{c}{2})(\annihilation{a}{}+\creation{a}{})$, with $\kappa\propto D_{12}$. The associated dissipator can be written as
\begin{eqnarray}
 \mathcal{D}'(\rho(t)) &=& \gamma_0\mathcal{P}_1^{(+)}\rho(t) \mathcal{P}_1^{(+)}-\frac{1}{2}\gamma_0\{\mathcal{P}_1^{(+)},\rho(t)\}\nonumber\\
 && + 4\gamma'_0 \mathcal{P}_2 \rho(t) \mathcal{P}_2-2 \gamma'_0\left\{\mathcal{P}_2,\rho(t)\right\},
\label{eq:dissipator local} 
\end{eqnarray}
where $\mathcal P_2 = \ket{e_1e_2}\bra{e_1e_2}$ is a Lindblad generator that induces dephasing of the doubly excited state. Therefore the two-molecule entangled state $\ket{\Phi} = a\ket{g_1g_2}+b\ket{e_1e_2}$ no longer belongs to a Decoherence-Free Subspace (DFS) with respect to the phonon environment, i.e. $\mathcal{D}'(\rho(t))\neq 0$. The decoherence rate $\gamma'_0$ would depend on the magnitude of the dipolar shift $D_{ij}$, which can be tuned by manipulating the strength of an applied static electric field and the intensity of the trapping laser. In addition to the phonon-induced fluctuations of the site energies in the presence of DC electric fields, the molecular energies also undergo fluctuations due to electric field noise, which acts as a global source of decoherence that can lead to entanglement decay as discussed for general bipartite and tripartite states in Refs. \cite{Yu:2003}.

\section{Model spectral density of optical lattice phonons}
\label{sec:spectral density}

In this appendix we derive the expression for the transition rate $\gamma_{\mu\nu,\mu'\nu'}(\omega)$ in Eq. (\ref{eq:transition rate}) using a semiclassical model for the phonon environment in optical lattices. We start from the system-bath interaction operator in the exciton basis $ \hat H_{SB} = \sum_{\mu\nu}\sum_k \lambda_{\mu\nu}^k\crea{c}{\mu}\anni{c}{\nu}\left(\anni{a}{k}+\crea{a}{k}\right)$
and define the time correlation function $C_{\mu\nu,\mu'\nu'}(t) = \langle \hat B_{\mu\nu}(t)\hat B_{\mu'\nu'}(0)\rangle$, where the bath operator $\hat B(t)$ in the interaction picture is given by $ \hat B_{\mu\nu}(t) = \sum_k \lambda_{\mu\nu}^k\left[\anni{a}{k}(t)+\crea{a}{k}(t)\right]$.

The classical vibrational energy of the array can be written as $H = (1/2)\sum_k\dot Q_k^2 +\omega_k^2 Q_k^2$, where $Q_k = \sum_{j=1}^\mathcal{N} \alpha_{jk} \sqrt{m}\,x_j$ are the normal modes of vibration defined in terms of the displacements $x_j$ from equilibrium and the molecular mass $m$. Promoting normal coordinates to quantum operators as $\hat Q_k = \sqrt{\hbar/2\omega_k}\left(\anni{a}{k}+\crea{a}{k}\right)$ allows us to write the semiclassical bath operator $ B^{\rm cl}_{\mu\nu}(t) = \sum_k \lambda_{\mu\nu}^k \sqrt{\frac{2\omega_k}{\hbar}}Q^{\rm cl}_k(t)$. 
The {\it classical} bath correlation function 
can thus be written as
\begin{equation}
 C_{\rm cl}(t) = \sum_k \lambda_{\mu\nu}^k\lambda^k_{\mu'\nu'}\left(\frac{2\omega_k}{\hbar}\right)\langle Q_k(t)Q_k(0)\rangle_{\rm cl},
 \label{app:classical TCF}
\end{equation}
where we used the fact that different modes ($k'\neq k$) are uncorrelated. The classical bath correlation function is a real quantity, i.e., $C^*_{\rm cl}(t) = C_{\rm cl}(t)$.

The quantum bath correlation function (omitting system state indices) is defined as $C(\tau) = \langle \hat B(\tau)\hat B(0)\rangle$ and satisfies $C^*(t) = C(-t)$ \cite{Breuer-Petruccione-book}. The system transition rate is given by $\gamma(\omega) = G(\omega)/\hbar^2$ where $G(\omega) = \int_{-\infty}^\infty d\tau \rme^{i\omega\tau} C(\tau)$ is a real positive quantity. Using the detailed balance condition $G(-\omega) = \rme^{-\beta\hbar\omega}G(\omega)$, where $\beta = 1/k_{\rm b}T$, it is possible to write
\begin{equation}
 G(\omega) = \frac{2}{1-\rme^{-\beta\hbar\omega}}G_A(\omega),
\end{equation}
where $G_A(\omega) = \int_{-\infty}^{\infty} d\tau \rme^{i\omega\tau}{\rm Im}\{C(\tau)\}$. We use this expression to obtain a semiclassical approximation to the quantum rate $\gamma(\omega)$.


 The approximation scheme consists on relating the antisymmetric function $G_A(\omega)$ to the Fourier transform $G_{\rm cl}(\omega)=\int_{-\infty}^\infty \rme^{i\omega\tau}C_{\rm cl}(\tau)$ of the classical bath correlation function in Eq. (\ref{app:classical TCF}). Following Ref. 
 \cite{Egorov:1999}, 
 we use $G_A(\omega)\approx (\beta\hbar\omega/2)\,G_R(\omega)$, and postulate the semiclassical closure $C_R(t) = C_{\rm cl}(t)$.
 This procedure is known as the harmonic approximation. The approximate quantum transition rate is thus given by
\begin{equation}
 \gamma(\omega) = \frac{1}{\hbar^2} \frac{\beta\hbar\omega}{1-\rme^{\beta\hbar\omega}}G_{\rm cl}(\omega).
 \label{app:semiclassical rate}
\end{equation}

The next step is specific to the system considered here. It involves the evaluation of the correlation function $\langle Q_k(t)Q_k(0)\rangle_{\rm cl}$ from the classical equations of motion of a molecule in the optical lattice potential. For simplicity, we consider the potential to have the harmonic form $V(x) = \frac{1}{2}m\omega_k^2x^2$, where $\omega_k$ is the frequency of the normal mode $k$. The most general form of the mode frequency is $\omega_k = \omega_0 f(k)$, where $\omega_0 = (2/\hbar)\sqrt{V_LE_R}$ is the trapping frequency as determined by the lattice depth $V_L$ and the recoil energy $E_R$ of the molecule. The function $f(k)$ accounts for the dispersion of the phonon spectrum and is determined by the dipole-dipole interaction between ground state molecules in different lattice sites \cite{Herrera:2011}. In this work we consider molecules in the absence of static electric fields, therefore the induced dipole moment vanishes and the phonon spectrum is dispersionless. For any $k$, the mode 
frequency $\omega_k = \omega_0$ thus depends on the trapping laser intensity $I_L$ since $V_L \propto I_L$ \cite{Bloch:2005,Carr:2009}. The laser intensity noise therefore modulates the phonon frequency $\omega_0$ and can lead to heating when the noise amplitude is large enough \cite{Savard:1997,Gehm:1998}. The motion of a molecule in a fluctuating harmonic potential can be modeled by the equation of motion (for each $k$)
\begin{equation}
 \ddot Q_k +\omega_k^2(t) Q_k = 0,
 \label{app:SEOM}
\end{equation}
where $\omega_k^2 = \omega_0^2\left[1+\alpha\xi(t)\right]$, and $\alpha\xi(t)$ is proportional to the relative intensity noise, i.e, $\alpha\xi(t) \propto (I_L(t)-\langle I_0\rangle)/\langle I_0\rangle$. 

The equation of motion in Eq. (\ref{app:SEOM}) is a stochastic differential equation with multiplicative noise, for which no exact analytical solution exists \cite{VanKampen-book}. 
Using a cumulant expansion approach, the equation of motion for the correlation function $\langle Q(t)Q(0)\rangle$ can be written as \cite{VanKampen-book}
\begin{equation}
\frac{d^2}{dt^2}\langle Q(t)Q(0)\rangle+2\beta\frac{d}{dt}\langle Q(t)Q(0)\rangle + \omega_0^{'2}\langle Q(t)Q(0)\rangle = 0,
\label{app:correlation EOM}
\end{equation}
where $\beta = \alpha^2\omega_0^2c_2/4$ is an effective noise-induced damping coefficient and $\omega_0^{'2} = \omega_0^2(1-\alpha^2\omega_0c_1)$ is an effective oscillator frequency which includes a noise-induced shift from the deterministic value $\omega_0$. Equation (\ref{app:correlation EOM}) is valid for all times provided $\alpha\tau_c\ll 1$, where $\tau_c$ is the noise autocorrelation time. The coefficients $c_1$ and $c_2$ are related to the noise autocorrelation function by 
\begin{eqnarray}
 c_1  &=& \int_0^\infty \langle \xi(t)\xi(t-\tau)\rangle \sin(2\omega_0\tau)d\tau\\
 c_2 &=& \int_0^\infty \langle \xi(t)\xi(t-\tau)\rangle [1-\cos(2\omega_0\tau)]d\tau.
\end{eqnarray}
The effective damping constant can thus be written as $\beta = (\alpha^2\omega_0^2/8)[S(0)-S(2\omega_0)]$, where $S(\omega) = \int_{-\infty}^\infty\langle \xi(t)\xi(t-\tau)\rangle\rme^{-i\omega\tau}d\tau$ is the noise spectral density. The dependence of the damping coefficient on the spectral density at twice the natural frequency indicates that this is parametric dynamical process that can lead to heating ($\beta<0$) when $S(2\omega_0)>S(0)$. Here we assume that the static laser noise is dominant and use $\beta>0$, which is satisfied for trapping lasers with approximate $1/f$ noise as in Ref. \cite{Savard:1997}. 

The solution to Eq. (\ref{app:correlation EOM}) is $\langle Q(t)Q(0)\rangle = \langle Q^2(0)\rangle\rme^{-\beta |t|}\cos(\omega' t)$, with $\omega' = \sqrt{\omega_0^2-\beta^2}$. We have assumed the oscillator is underdamped ($\omega_0>\beta$), and ignored the noise-induced frequency shift ($\omega_0' = \omega_0$). The mean square amplitude $\langle Q^2(0)\rangle$ can be obtained by averaging over initial conditions using Boltzmann statistics. For an ensemble of identical one-dimensional harmonic oscillators we have $\langle Q^2(0)\rangle = k_{\rm b}T/\omega_0^2$. Combining these results we can write the classical bath correlation function in Eq. (\ref{app:classical TCF}) as
\begin{equation}
 C_{\rm cl}(t) = \sum_k\lambda_{\mu\nu}^k\lambda_{\mu'\nu'}^k\left(\frac{k_{\rm b}T}{\hbar\omega_k}\right)\rme^{-\beta|t|}\cos(\omega'_kt).
 \label{app:classical TCF thermal}
\end{equation}
By inserting the Fourier transform of Eq. (\ref{app:classical TCF thermal}) into Eq. (\ref{app:semiclassical rate}) we obtain the semiclassical transition rate
\begin{equation}
 \gamma_{\mu\nu,\mu'\nu'}(\omega) = \frac{1}{\hbar^2}\left[n(\omega)+1\right]\left[J^{\rm cl}_{\mu\nu,\mu'\nu'}(\omega)-J^{\rm cl}_{\mu\nu,\mu'\nu'}(-\omega)\right],
 \label{app:rate vs SD}
\end{equation}
%
%
where $n(\omega) = (\rme^{\beta\hbar\omega}-1)^{-1}$ is the Bose distribution function and we have defined the semiclassical phonon spectral density 
\begin{equation}
 J^{\rm cl}_{\mu\nu,\mu'\nu'}(\omega) = \sum_k\lambda_{\mu\nu}^k\lambda_{\mu'\nu'}^k\left(\frac{\omega}{\omega_k}\right)\frac{\beta}{(\omega-\omega_k')^2+\beta^2}.
 \label{app:spectral density}
\end{equation}
This approximate expression for $J(\omega)$ should be compared with exact phonon spectral density for an ensemble of free quantum oscillators $J_{\mu\nu,\mu'\nu'}(\omega) = \omega^2\sum_k\lambda_{\mu\nu}^k\lambda_{\mu'\nu'}^k\delta(\omega-\omega_k)$, which also satisfies Eq. (\ref{app:rate vs SD}).

\section*{References}

\bibliographystyle{unsrt}
\bibliography{ame-v2}

\end{document}